\documentclass[12pt,a4paper,final]{iopart}
\usepackage{citesort}
\usepackage[dvipsnames]{xcolor}
\usepackage{graphicx}% Include figure files
\usepackage{geometry}
\usepackage{caption}
\usepackage{subcaption}
\usepackage{tabularx}
\usepackage{xfrac}
\usepackage{relsize}
\usepackage{scalerel}
\usepackage{esvect} 

\newcolumntype{L}{>{$}l<{$}}
\newcommand{\be}{\begin{equation}}
\newcommand{\bea}{\begin{align}}
\newcommand{\ee}{\end{equation}}
\newcommand{\eea}{\end{align}}
\usepackage{array} 
\begin{document}
\nocite{*}
\title[]{Comparing the roles of time overhead and spatial dimensions on optimal resetting rate vanishing transitions, in Brownian processes with potential bias and stochastic resetting}
\author{Saeed Ahmad, and Dibyendu Das}
\address{Physics Department, Indian Institute of Technology Bombay, Mumbai 400076, India}
\ead{saeedmalik@iitb.ac.in}
\vspace{10pt}
\begin{indented}
\item[]Dec 2022
\end{indented}
\begin{abstract}
  The strategy of stochastic resetting is known to expedite the first passage to a target, in diffusive systems.  Consequently, the mean first passage time is minimized at an optimal resetting parameter. With Poisson resetting, vanishing transitions in the optimal resetting rate $r_*\to 0$ (continuously or discontinuously) have been found for various model systems, under the further influence of an external potential. In this paper, we explore how the discontinuous transitions in this context are regulated by two other factors --- the first one is spatial dimensions within which the diffusion is embedded, and the second is the time delay introduced before restart. We investigate the effect of these factors by studying analytically two types of models, one with an ordinary drift and another with a barrier,  and both types having two absorbing boundaries. In the case of barrier potential, we find that the effect of increasing dimensions on the transitions is quite the opposite of increasing refractory period.
\end{abstract}
 
\section{\label{sec:sec1}Introduction}
First passage processes with stochastic resetting have been widely studied, in the past years \cite{PRL_Evan_Staya_2011,J_Phys_A_Evan_Staya_2011Position,J_Phys_A_Evans_Non_equil_2013,J_Phys_A_math_cogu_diff_w_r_Durang_2014,Ref_1_PRL_Satya_Sanjib_First_2014,J_Phys_A_Evans_Staya_higher_d_2014,J_Stat_Mech_determinitic_Bhat_2016,PRE_pwer_law_rate_Apoorva_ShamikG_2016,J_Phys_A_ApalKundu_Evan2016reset_t,PRL_Staya_fluctuating_interface_2014,Arxiv_RK_singh_Probability_Relax,Review_Evans_Satya_Reset_application_2019,PRL_Reuveni_2016First,PRL_Montanari_2002,PRE_Apal_descrete_time_renewal_2021,PNAS_Reuveni_Shlomi_2014_MicMenten,PRE_Sajib_2015Morkov_reset,Ref_2_PRE_Lukas_Ewa_2015,Ref_3_PRL_Belan_2018,PRR_Ising_Model_w_reset_Majumdar_2020,Ref_4_PRR_Reuveni_home_return_2020,PhysRevE_Bressloff_2020,PRE_Staya_Sanjib_2015_temporal_relaxation,J_Phys_A_accele_reset_Singh_2020,PRE_Bodrova_D(t)_reset_2019,PRE_Telegrap_w_reset_2018,PRL_Optimization_w_reset_FP_2020,PRR_Madian_mode_FP_w_reset_2020,PRE_Non_inst_reset_Badrova_2020,J_Phys_Commun_Susceptivility_reset_Grange_2020,PRE_Extereme_Statistics_w_reset_2021,J_Phys_A_non_inst_reset_linear_pot_Gupta_2021,J_Stat_Mech_Reset_under_damped_Gupta_2019,PRE_Asymmetric_reset_D_Gupta_2020,Soft_matter_Reset_w_Lorentz_force_2021,PRL_Reset_w_branching_Reuveni_2020}. It is a strategy that interrupts the process in between, only to restart again. This simple mechanism produces two major implications on ordinary diffusive processes. On one hand, it can produce a non-equilibrium stationary state \cite{PRL_Evan_Staya_2011,J_Phys_A_Evans_Staya_higher_d_2014,J_Phys_A_math_cogu_diff_w_r_Durang_2014,PRL_Staya_fluctuating_interface_2014,PRE_APal_2015Potential,J_Phys_A_reset_confining_pot_Metzler_Singh_2020,Arxiv_RK_singh_Probability_Relax}  and on the other hand, the mean first passage time (MFPT) to reach a target is optimized \cite{PRL_Evan_Staya_2011,J_Phys_A_Evans_Staya_higher_d_2014}. Its significant applications have been found in many branches of science \cite{Review_Evans_Satya_Reset_application_2019,Brief_Review_reset_Shamik_G_2022}. Various problems have been studied in biological contexts \cite{PRE_backtrac_RNA_polymer_Roldan_2016,J_Phys_A_Run_and_tumble_Evans_Staya_2018,J_Phys_A_Run_and_tumble_w_reset_2d_Santra_2020,PRE_mutiple_target_Bressloff_2020,EPL_Antiviral_reset_Ramoso_2020,J_Phys_A_partial_ab_Bresloff_Schumm_2021}, in the context of chemical reactions \cite{PNAS_Reuveni_Shlomi_2014_MicMenten,PRE_Reuveni_Shlomi2015_MicMenten,Nature_Single_Enzyme_Robin_S_Reuveni_2018,PRE_Optimal_reset_polination_2019} and computer science \cite{PRL_Montanari_2002,Compt_sys_Optimal_reset_Lorenz_2021}. The resetting event may be instantaneous \cite{PRL_Evan_Staya_2011} or non-instantaneous \cite{PRE_Non_inst_reset_Badrova_2020,PRE_Non_inst_reset_Pal_Ku_Reuveni_2019,New_Jour_Phys_Non_inst_reset_Pal_2019}, while time intervals between resets may be exponentially distributed \cite{PRL_Reuveni_2016First,J_Stat_Mech_determinitic_Bhat_2016}, power-law distributed \cite{PRE_pwer_law_rate_Apoorva_ShamikG_2016,PRL_APAl_Reuveni_2017FirstR}, or fixed \cite{J_Stat_Mech_determinitic_Bhat_2016,PRL_APAl_Reuveni_2017FirstR}. Recently resetting of a bead diffusing on fluid have been studied experimentally using optical tweezers \cite{J_Chem_letter_Reuveni_experimental-reset_2020} and optical traps \cite{PRR_Experiment_evidence_Majumdar_2020}. 

In this paper we focus on the Poisson resetting process, with resetting events happening at a constant rate ~\cite{PRL_Evan_Staya_2011,PRL_Reuveni_2016First,PRL_APAl_Reuveni_2017FirstR,Review_Evans_Satya_Reset_application_2019}. For such cases, it has been shown that the mean first passage time (MFPT) to reach a target is minimized at an optimal resetting rate $r_*$ (ORR), for which the coefficient of variation (CV) (i.e. the ratio of standard deviation to mean) of the process is unity \cite{PRL_Reuveni_2016First}. Introducing a potential bias towards the target always expedite the first passage. The simultaneous presence of both resetting and potential bias has been studied in different models \cite{PRE_APal_2015Potential,PRE_Saeed_2019,J_Phys_A_Ray_Debasish_Reuveni_2019,PRR_Pal_Parsad_Landau_2019,J_Chem_Phys_Reuveni_log_Pot_2020,PRE_Saeed_Das_2020,J_Phys_A_non_inst_reset_linear_pot_Gupta_2021,J_Chem_Phys_Reuveni_Ray_dble_w_2021,PRE_Huang_higher_d_reset_2_ab_2022,PRE_Saeed_2022}. It is generically found that the ORR  is strongly regulated by the  strength of the potential and the reset location. In particular, beyond some threshold potential strength and initial (and reset) position threshold, resetting becomes disadvantageous. Thus, the ORR undergoes a dynamical transition ($r_*\to 0^{+}$) from a resetting beneficial state to a state where the strategy hinders~\cite{J_Phys_A_Christo_Reset_Boun_2015,PRE_Cristou_reset_circle_2018,PRE_Saeed_2019,J_Phys_A_Ray_Debasish_Reuveni_2019,PRR_Pal_Parsad_Landau_2019,J_Chem_Phys_Reuveni_log_Pot_2020,J_Chem_Phys_Reuveni_Ray_dble_w_2021}. Study of such  transitions have been extended to dimensions greater than one ~\cite{PRE_Saeed_Das_2020,PRE_Huang_higher_d_reset_2_ab_2022}.

In general the ORR transition may be continuous \cite{PRE_Saeed_2019,J_Phys_A_Ray_Debasish_Reuveni_2019,PRR_Pal_Parsad_Landau_2019,J_Chem_Phys_Reuveni_log_Pot_2020,PRE_Saeed_Das_2020,J_Chem_Phys_Reuveni_Ray_dble_w_2021,PRE_Huang_higher_d_reset_2_ab_2022,PRE_Saeed_2022} as well as discontinuous \cite{Ref_1_PRL_Satya_Sanjib_First_2014,Ref_5_PRE_2015,PRR_Pal_Parsad_Landau_2019,PRE_Huang_higher_d_reset_2_ab_2022,PRE_Saeed_2022}  in nature. A Landau-like description has been recently developed \cite{PRE_Saeed_2019,PRR_Pal_Parsad_Landau_2019}. From the latter, the exact location of a continuous transition and tri-critical point \cite{PRR_Pal_Parsad_Landau_2019}, if present, may be found from the knowledge of cumulants of first passage times without resetting, for any system in general. There is no such corresponding exact criterion for a discontinuous transition though. The Landau theory gives  approximate results for both jump size of ORR and the location of its transition, in terms of cumulants, but they become unreliable when the discontinuities in ORR are large. In the absence of  a general criterion for discontinuous transitions,  specific case studies  under diverse circumstances teach us selective lessons about when such transitions may arise. In this article, we pursue this general idea of understanding the discontinuous ORR transitions through analytically solvable models.

Analytical study of certain model systems show that discontinuous transition in ORR seems to arise in systems with two targets for first passage and some asymmetry which may be due to a potential  \cite{Ref_5_PRE_2015,PRR_Pal_Parsad_Landau_2019,PRE_Huang_higher_d_reset_2_ab_2022,PRE_Saeed_2022}. On the other hand,  a model with Levy flights and resetting seems to be an exception where discontinuous transition happens even with a single target  \cite{Ref_1_PRL_Satya_Sanjib_First_2014}. In a recent study, where resetting is done using an optical trap to variable locations spread over a finite spatial range, the finite ORR at a metastable minimum of MFPT jumps to infinity, if the spatial range is tuned above a critical value \cite{PRR_Experiment_evidence_Majumdar_2020}. Another recent study suggests that spatial dimension is also important.  For example, in the problem with diffusion in finite confinement with two absorbing boundaries, under resetting, the transition in ORR is continuous  in $d=1$, but discontinuous  for $d>1$ \cite{PRE_Huang_higher_d_reset_2_ab_2022}. We observe an interesting fact in this context --- while in $d=1$ two boundaries of a finite domain are equivalent (points), in  $d>1$ the two spherical boundaries of any finite domain have unequal areas --- smaller for the inner boundary and larger for the outer boundary.  This inherent geometrical asymmetry in $d>1$  \cite{PRE_Huang_higher_d_reset_2_ab_2022} seems like a parallel to having an asymmetry through potential bias in $d=1$ with two absorbing boundaries   \cite{PRR_Pal_Parsad_Landau_2019,PRE_Saeed_2022}, where discontinuous  transition in ORR is seen. Another factor that may play a role is introduction of time delay before restart \cite{PRL_Reuveni_2016First,PNAS_Reuveni_Shlomi_2014_MicMenten,PRE_Reuveni_Shlomi2015_MicMenten,Nature_Single_Enzyme_Robin_S_Reuveni_2018,J_Phys_A_Refractory_Period_w_reset_Evans_2019}.  The effect of time overhead $\langle T_{on} \rangle$ on continuous transitions in ORR for biased diffusive processes have been analyzed in $d \geq 1$ \cite{PRE_Saeed_2019,PRE_Saeed_Das_2020} showing lowering of the critical bias strength, but its effect on discontinuous transitions is relatively unexplored. In enzymatic reactions, $\langle T_{on}\rangle$ is the binding time of an enzyme to substrate, which is inverse of binding rate $k_{on}$ \cite{PNAS_Reuveni_Shlomi_2014_MicMenten,PRE_Reuveni_Shlomi2015_MicMenten}.
%which shows that for non-zero delay transition happen at the lower values of critical strength.

In this article we study the effect of spatial dimensions, and time delay before restart, on the ORR transitions. We explore whether these two factors have opposite or similar effect.  Moreover we study this within models in confined space which either have a simple drift, or a barrier potential. We have shown recently that in $d=1$, that rather interesting phase transitions  are associated with the barrier potentials -- with one absorbing boundary, transitions are continuous but the available space breaks up into multiple resetting beneficial regions, while with two boundaries there are multiple tri-critical points and exotic `finite' lines of discontinuous transition \cite{PRE_Saeed_2022}. The latter paper was motivated by the question of resetting in a magnetic system at low temperature where the system may toggle between two magnetization states with a potential barrier in between. The diffusion was interpreted to occur in a space of scalar order parameter. Extension of the problem to higher dimensions ($d>1$) would be necessary if the order parameter is a vector with multiple components.  What role does resetting  play in higher dimensions in problems with barriers are thus an important theoretical question. The exact MFPT under resetting has been found for unbiased diffusion in semi-infinite domain \cite{J_Phys_A_Evans_Staya_higher_d_2014}, and for logarithmic and box potentials \cite{PRE_Saeed_Das_2020} in any general dimension $d$. For a linear potential (i.e. ordinary drift), the exact MFPT is known in $d=1$ \cite{PRE_Saeed_2019,J_Phys_A_Ray_Debasish_Reuveni_2019}, but the explicit solution of MFPT in higher dimensions ($d > 1$) and with an average time delay $\langle T_{on} \rangle$ seems to be lacking. In this work we derive the later, and use it to obtain the continuous and  discontinuous transitions and the tri-critical points in the models with drift or barrier potentials in confined spaces. For the barrier potential, we use piece-wise linear forms as was done in \cite{J_Phys_A_reset_confining_pot_Metzler_Singh_2020,PRE_Saeed_2022}. 

The article is organized as follows. In section \ref{sec:sec2}, we discuss the models of biased Brownian particle with stochastic resetting in $d$ dimensional space and finite refractory period $\langle T_{on} \rangle$, and also discuss a few relevant theoretical aspects. In section \ref{sec:sec3}, we obtain the analytical solution for the linear potential.  We use the latter result in sections \ref{sec:sec4} and \ref{sec:sec5}, to study the dynamical transitions of ORR for the simple drift potential and the barrier potential, respectively, by explicitly varying $\langle T_{on} \rangle$ and $d$. Finally, we conclude in section \ref{sec:sec6}.

\section{\label{sec:sec2} The problems of biased diffusion with resetting and time overhead in arbitrary spatial dimensions}
\subsection{The Models}
We consider a particle doing Brownian motion under a spherically symmetric potential $V(|\vec{R}|)$, in arbitrary spatial dimension $d$, within a finite domain bounded by two absorbing boundaries  at $|\vec{R}| = |\vec{a}|$ and $|\vec{R}| = |\vec{R}_m|$. The particle starts uniformly from a surface of radius $|\vec{R}|\in(|\vec{a}|,|\vec{R_m}|)$ at time $t=0$, and is repeatedly stochastically reset to a point chosen uniformly on a spherical surface of radius $|\vec{R}_0|$ (distinct from initial radius $|\vec{R}|$) at a constant rate $r$, unless it attains first passage at either of the two absorbing boundaries.  Moreover between a reset and a fresh restart of the motion, there is a time delay period $\langle T_{on} \rangle$. 

Rather than instantaneous restart after reset, having a finite refractory period $\langle T_{on} \rangle$ is more natural and has been considered in many contexts including enzymatic reactions \cite{PRL_Reuveni_2016First,PNAS_Reuveni_Shlomi_2014_MicMenten,PRE_Reuveni_Shlomi2015_MicMenten,Nature_Single_Enzyme_Robin_S_Reuveni_2018,J_Phys_A_Refractory_Period_w_reset_Evans_2019}. Typically MFPT increases in the presence of nonzero $\langle T_{on} \rangle$ \cite{PNAS_Reuveni_Shlomi_2014_MicMenten,PRE_Reuveni_Shlomi2015_MicMenten}.  When potential biases interplay with stochastic resetting, specific studies have shown that the critical potential strength at ORR transition is lowered for a finite  $\langle T_{on} \rangle$ \cite{PRE_Saeed_2019,PRE_Saeed_Das_2020} --- thus finite refractory period lowers the importance of resetting in the presence of potential bias. On the other hand, rising spatial dimension $d$ enhances the possible directions of diffusion and hence fluctuations, and consequently resetting becomes increasingly important. To offset resetting and have a continuous ORR transition, higher potential strength is thus needed in higher $d$ \cite{PRE_Saeed_Das_2020}.  
%%%%%%%%%%%%%%%%%%%%%%%%
%%%%%%%%%%%%%%%%%%%%%%%%
%%%%%%%%%%%%%%%%%%%%%%%%
%\iffalse
\begin{figure}[ht!]
 \begin{subfigure}{0.475\textwidth}
 \includegraphics[width = 1\textwidth,height=0.18\textheight]{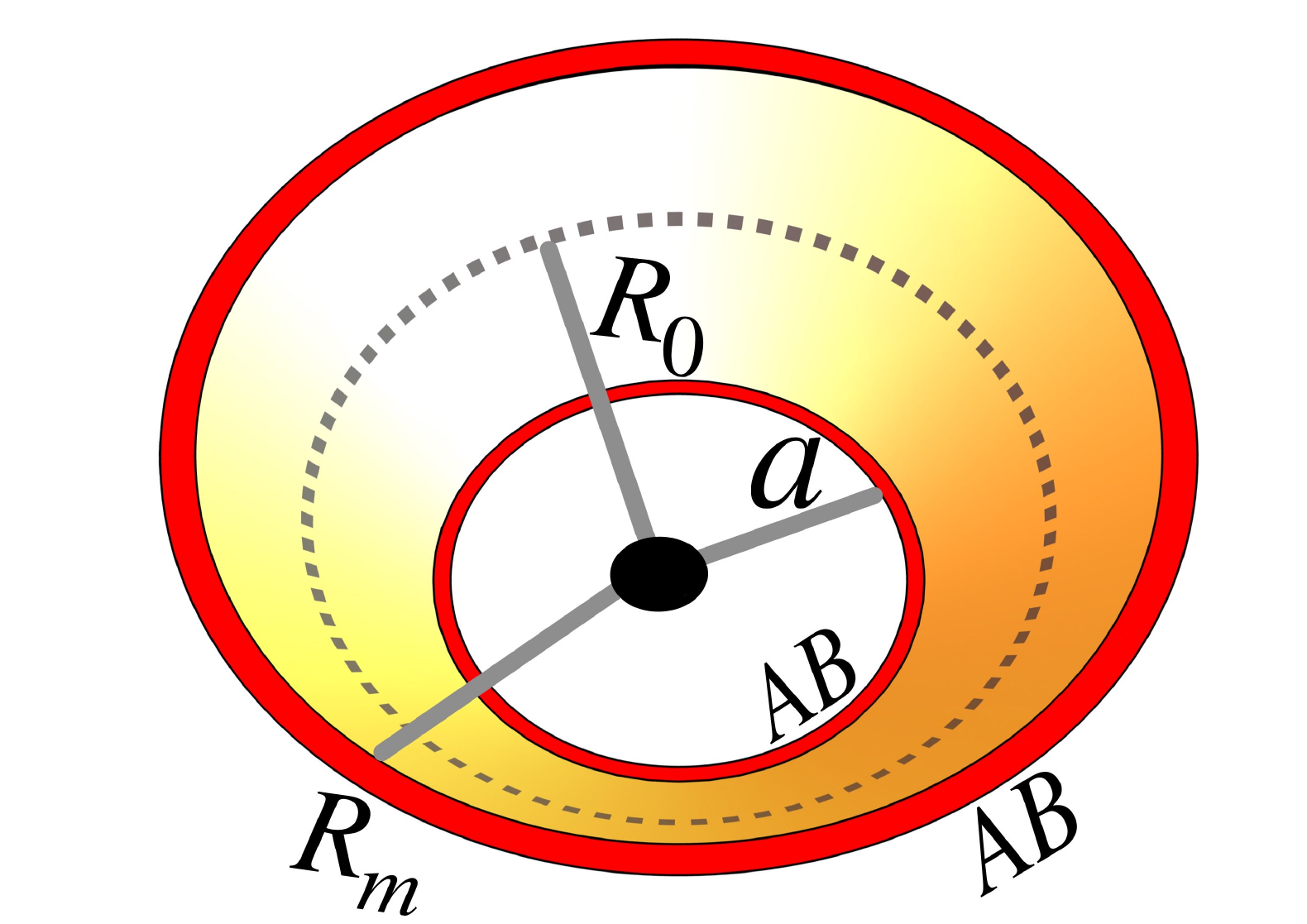}
   \caption[]{} 
   \label{fig:d_dim_linear_pot}
  \end{subfigure}
\begin{subfigure}{0.475\textwidth}
    \includegraphics[width = 1\textwidth,height=0.18\textheight]{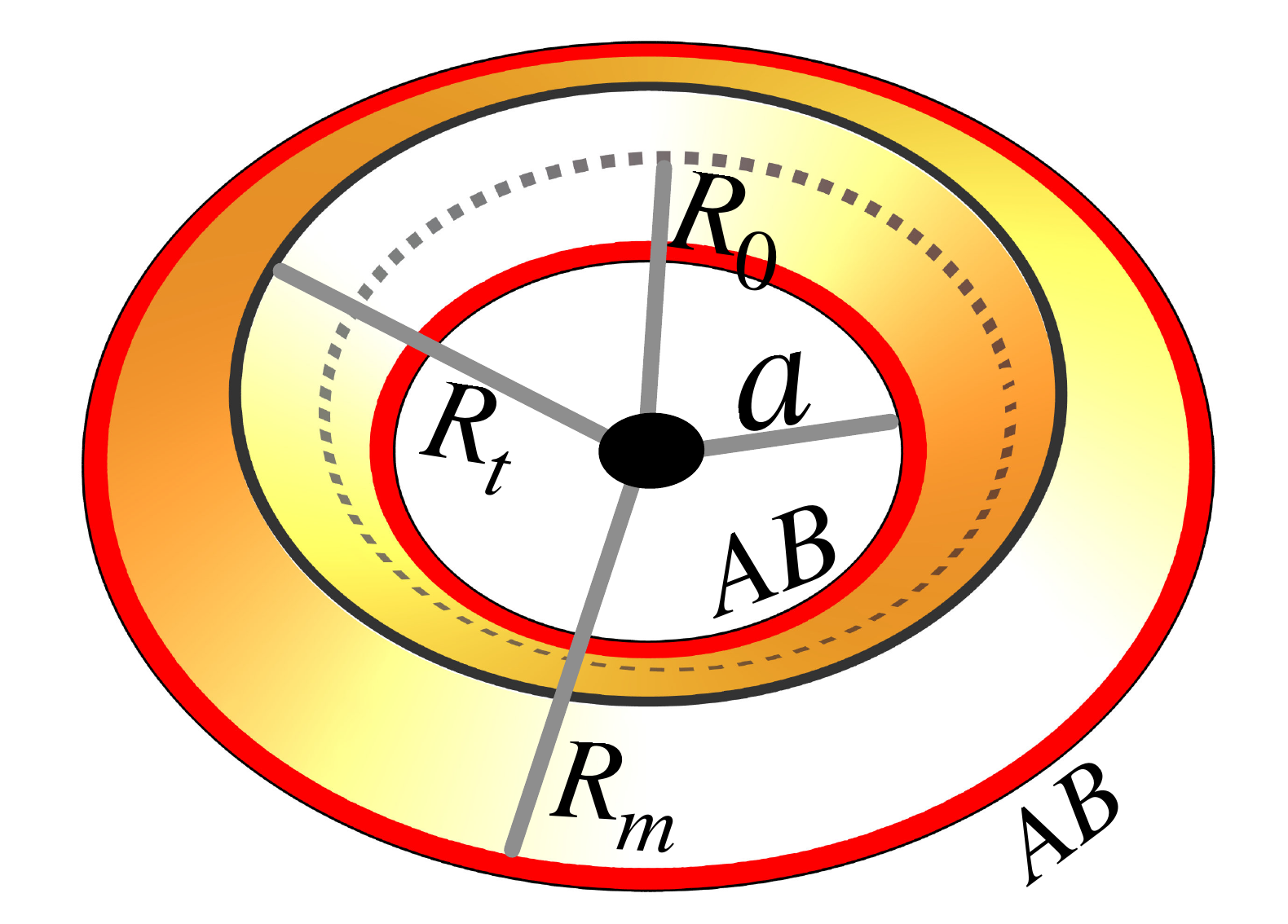}
    \caption[]{} 
   \label{fig:d_dim_pw_linear_ordr}
 \end{subfigure}
  \caption{Radially symmetric potentials over two-dimensional space are shown between two absorbing boundaries (AB) at inner radius $a$ and outer radius $R_m$, for two cases: (a) linear and (b) piece-wise linear with a peak at $R_t$ and valleys at the two boundaries. The reset radius is $R_0$ in both cases.  
  }
  \label{fig:the_models}
\end{figure}
%\fi
%%%%%%%%%%%%%%%%%%%%%%%%
%%%%%%%%%%%%%%%%%%%%%%%%
%%%%%%%%%%%%%%%%%%%%%%%%

In this study we consider potentials of two different types, between the two absorbing targets at $|\vec{a}|$ and $|\vec{R}_m|$. The first one is the linear potential 
\be
V(R)=k R, ~~ a \leq R \leq  R_m 
\label{lin_pot}
\ee
having a positive slope (i.e. $k > 0$) which produces a drift towards the inner radius $a$. Although visual depiction in dimensions $d>2$ is difficult, we show it for $d=2$ in Fig.(\ref{fig:d_dim_linear_pot}). The second potential (shown in Fig. (\ref{fig:d_dim_pw_linear_ordr}) for $d=2$) is a  piece-wise linear potential with a barrier, mathematically defined as 
\be
V(R) =
\left\{\begin{array}{@{\kern2.5pt}lL}
    \hfill k_1(R-R_t), & $a \leq R\leq R_t$,\\
    \hfill -k_2(R-R_t), & $R_t \leq R\leq R_m$.
\end{array}\right.
\label{Tent_pot}
\ee
Here if $k_1 > 0$, then the drift in the region $\in [a,R_t)$ is towards the inner boundary at $a$, while if $k_2 > 0$, there is a similar drift towards the outer boundary $R_m$ in the region $\in (R_t, R_m]$.  In our study $k_2$ is varied over positive and negative values. An interesting fact to observe is that, unlike this model studied earlier in one-dimension \cite{PRE_Saeed_2022}, the volumes of the spaces in the  two regions $a<R<R_t$ and $R_t<R<R_m$ are unequal as are the surface areas of absorption at $a$ and $R_m$.  Thus even if $k_1 = k_2$, the potential bias on the two sides of $R_t$ are not the same --- this is an unavoidable asymmetry in any $d>1$.

\subsection{The theoretical problem of First Passage and relevant quantities}
We are interested to obtain the MFPT  $\langle T_{r} \rangle$ in the presence of resetting in the various models, and the behavior of the 
resetting rate $r_*$ which optimizes $\langle T_{r} \rangle$, as a function of other parameters of the model. Using renewal theory, an 
important relationship between the MFPT and the Laplace transform of distribution of the first passage time without resetting namely $\tilde{F}(s)$ has been derived \cite{PRL_Reuveni_2016First}:
\begin{equation}
  \langle T_{r} \rangle =\frac{\langle T_{on} \rangle}{\tilde{F}(r)}+\frac{1-\tilde{F}(r)}{r\tilde{F}(r)}.
  \label{eq:general_MFPT}
 \end{equation}
 Thus if $\tilde{F}(s)$ is known, then for a given problem the MFPT and the associated transitions in ORR $r_*$ may be studied. It is also known that  $\tilde{F}(s)$ is related to the Laplace transform of the survival probability $q(R,s)$ (see \cite{redner2001guide}):
\be
\tilde{F}(R,s)=1-sq(R,s).
\label{eq:FPT_Survival}
\ee
%In the following we describe the method to find the transition in ORR with  $\langle T_{on} \rangle$.
%\subsection{\label{subsec:sec21}Landau like description for stochastic time overhead}

We are interested in transitions in ORR $r_*$ where it vanishes continuously or discontinuously as a function of parameters of the model. Near such a transition, $r_*$ may be assumed to be small. This is indeed true for continuous transitions and tri-critical point, but may not be a good approximation for a discontinuous transition if the jump in $r_*$ is large. Under small $r$ assumption, a Taylor series expansion of the above Eq.(\ref{eq:general_MFPT}) leads to
\be
\langle T_r \rangle= a_0+a_1r+a_2r^2+a_3r^3+\mathcal{O}(r^4),
\label{eq:small_r_exp}
\ee
where $a_0=\langle T \rangle+\langle T_{on} \rangle$, $a_1=\langle T \rangle^2+\langle T \rangle\langle T_{on} \rangle-\frac{\langle T^2 \rangle}{2}$, $a_2=\frac{\langle T^3 \rangle}{3!}+\langle T \rangle^3-\langle T \rangle \langle T^2 \rangle+\langle T_{on} \rangle\big(\langle T \rangle^2-\frac{\langle T^2 \rangle}{2} \big)$. In this small series expansion the sign of $a_0$ and $a_3$ must be non-negative. The detailed discussion of this may be found in \cite{PRR_Pal_Parsad_Landau_2019,PRE_Saeed_2022} --- note that here $\langle T_{on}\rangle \neq 0$. The transition in ORR may happen as a function of any parameter in the model --- the strength of the potential (e.g. $k$, $k_1$, $k_2$), or the initial location $R_0$, which we represent by a generic symbol $\lambda$.   The continuous transition happens by tuning the controlled parameter $\lambda\to \lambda_c$ such that $a_1\to 0$ while $a_2>0$. In terms of $\sigma^2=\langle T^2\rangle-\langle T\rangle^2$ this implies
\be
 \text{At} \hskip0.05\textwidth \lambda_{c}:\hskip0.02\textwidth
\left\{\begin{array}{@{\kern2.5pt}lL}
    \sigma^2=\langle T\rangle^2 +2\langle T\rangle \langle T_{on}\rangle, & and,\\
    \langle T^3\rangle>6\langle T\rangle(\sigma^2+\langle T_{on}\rangle^2).
\end{array}\right.
\label{Conti_Trans}
\ee
 At a tri-critical point $\lambda_{tc}$, the coefficients  $a_1\to 0^{-}$ and $a_2\to 0^{+}$. This implies: 
\be
 \text{At} \hskip0.05\textwidth \lambda_{tc}:\hskip0.02\textwidth
\left\{\begin{array}{@{\kern2.5pt}lL}
    \sigma^2=\langle T\rangle^2 +2\langle T\rangle \langle T_{on}\rangle, & and,\\
    \langle T^3\rangle=6\langle T\rangle(\sigma^2+\langle T_{on}\rangle^2).
\end{array}\right.
\label{Trans_TCP}
\ee
Thus to study a continuous transition or a tri-critical point exactly, we need to find the solutions of moments without resetting, namely $\langle T\rangle$, $\langle T^2\rangle$ and $\langle T^3\rangle$. For the discontinuous transition this theory is not accurate, as one moves away from the tri-critical point. Thus, we do not rely on Landau like description for the discontinuous transition. To study the discontinuous transition in general, we use the expression of exact MFPT with resetting (Eq.(\ref{eq:general_MFPT})). 
%It is also true that we do not require to find the discontinuous transitions so long as, it is in between two consecutive tri-critical points. 

\section{\label{sec:sec3}Analytical solution of the first passage distribution and the moments for the linear potential in $d$ dimensions}
         As discussed above, from Eq.(\ref{eq:general_MFPT}) the MFPT for a problem with resetting may be solved if one knows the distribution of first passage time in Laplace space $\tilde{F}(R,s)$ `without resetting'. Since both the models (Eqs. (\ref{lin_pot}) and (\ref{Tent_pot})) have linear or piece-wise linear potentials, we first derive the solution of $\tilde{F}(R,s)$ for the linear potential $V = k R$ in $d$ dimensions. Note that from Eq. (\ref{eq:FPT_Survival}), the first passage distribution is related to the Laplace transform of the survival probability $q(R,s) = \int_0^{\infty} dt Q(R,t) e^{-st}$, where $Q(|\vec{R}|,t)$ denotes the probability that the particle has not reached the target(s) up to time $t$, starting from $|\vec{R}|$, in the `absence of resetting'. The derivative $V^{'}(R) = k$,  and the spherical symmetry of the potential and the absorbing boundaries imply that  the backward Fokker-Planck equation for the survival probability may be written in terms of the radial coordinate as (see \ref{App_sec11} for detailed discussion): 
\begin{equation}
\hspace{-0.4cm}\frac{\partial Q(R,t)}{\partial t}=D\frac{\partial^2 Q(R,t)}{\partial R^2}+\bigg(\frac{D(d-1)}{R}-k\bigg) \frac{\partial Q(R,t)}{\partial R}.
\label{eq_BFPE_1}
\end{equation}
For the above, the initial condition is $Q(R,0)=1$. 
The corresponding equation for the Laplace transform $q(R,s)$ is
%In the Laplace space the survival probability $q\equiv q(R,s)=\int^{\infty} _0e^{-st}Q(R,t)dt$ satisfies:
\begin{equation}
\frac{\partial^2 q}{\partial R^2}+\bigg(\frac{d-1}{R}-\frac{k}{D}\bigg) \frac{\partial q}{\partial R}-\alpha^2q=-\frac{1}{D}
\label{eq_BFPE_2}, 
\end{equation}
where $\alpha =\sqrt{\frac{s}{D}}$. 
The Eq.(\ref{eq_BFPE_2}) leads to the following homogeneous differential equation for $\tilde{F}(R,s)$:
\begin{equation}
\frac{\partial^2 \tilde{F}}{\partial R^2}+\bigg(\frac{d-1}{R}-\frac{k}{D}\bigg) \frac{\partial \tilde{F}}{\partial R}-\alpha^2\tilde{F}=0.
\label{eq_BFPE_3}
\end{equation}
 We have $\tilde{F}=1$ at the absorbing boundaries $R=a$ or $R=R_m$  as $q=0$ at these locations. The Eq. (\ref{eq_BFPE_3}) may be solved by using both numerical and analytical methods. Here we obtain $\tilde{F}(R,s)$ analytically.
%\par\texttt{\bf Solution of $\tilde{F}$ for linear segment:} 

The Eq.(\ref{eq_BFPE_3}) may lead to the following equation for $g(R,s)$, using an interesting transformation $\tilde F= e^{\phi R}g$: 
\begin{equation}
  R\frac{d^2 g}{d R^2}+\bigg((d-1)+(2\phi-\frac{k}{D})R\bigg)\frac{d g}{dR}+\bigg(\phi(d-1)+\big(\phi^2-\alpha^{2}-\frac{\phi k}{D}\big)R\bigg)g=0,
  \label{eq:g_diffeent_eq}
\end{equation}
Choosing the free parameter $\phi$ such that $\phi^2-\alpha^{2}-\frac{\phi k}{D}=0$ and scaling the position as $cR=R^{'}$, the differential equation (Eq. (\ref{eq:g_diffeent_eq})) converts into the well known  confluent hypergeometric differential equation \cite{simmons2016differential}. Thus, setting $\phi=\frac{1}{2}\bigg(\frac{k}{D}\pm\sqrt{\frac{k^2}{D^2}+4\alpha^2}\bigg)$ we get:
\begin{equation}
 R^{'}\frac{d^2 g}{d R^{'2}}+\bigg(\gamma-R^{'}\bigg)\frac{d g}{dR^{'}}-\theta g=0,
\end{equation}
where $\gamma=d-1$ and $c=\mp\sqrt{\frac{k^2}{D^2}+4\alpha^2}$ with $\theta=-\frac{\phi(d-1)}{c}$. The general solution of confluent hypergeometric differential equation is:
\begin{equation}
g=A\, _1F_1(\theta ;\gamma ;c R)+BU(\theta,\gamma,cR) 
\end{equation}
where  $\, _1F_1(\theta ;\gamma ;c R)$ is the Kummer function \cite{simmons2016differential}. Here $\gamma=d-1$ is a positive integer, and in such cases,  the second solution (Tricomi confluent hypergeometric function) has logarithmic factors as follows \cite{Mthemetica_HypergeometricU_Series}:
\hspace*{-2.0cm}\vbox{
\begin{eqnarray}
  U(\theta, \gamma,cR)=&\frac{(-1)^{\gamma}}{\Gamma(\theta - \gamma+1)}\bigg(\frac{\ln(cR)}{(\gamma-1)!}\, _1F_1(\theta ;\gamma ;c R)\nonumber\\
  &+\sum^{\infty}_{l=0}\frac{(\theta)_l(\psi(\lambda+l)-\psi(l+1)-\psi(l+\gamma))(cR)^l}{(l+\gamma-1)!l!}\nonumber\\
  &-\sum^{\gamma-1}_{l=1}\frac{(l-1)!(cR)^{-l}}{(1-\theta)_l(\gamma-l-1)!}\bigg)
  \end{eqnarray}
}

where $(\theta)_l=\frac{\Gamma(\theta+l)}{\Gamma(\theta)}$ is Pochhammer symbol and $\psi(l)=\frac{d\ln\Gamma(l)}{dl}$ is PolyGamma function \cite{Arfken_7ed_book}. Thus we have the solution of $\tilde{F}=e^{\phi R}g$ to be:
\begin{equation}
  \tilde{F}(R,s) =Ae^{\phi R}\, _1F_1(\theta ;\gamma ;c R)+Be^{\phi R}U(\theta,\gamma,cR)
  \label{eq:tilde_y_soln}
\end{equation}
where $c=\sqrt{\frac{k^2}{D^2}+4\alpha^2}$, $\gamma =d-1$, $\theta=-\frac{\phi(d-1)}{c}$ and $\phi=\frac{1}{2}\bigg(\frac{k}{D}-\sqrt{\frac{k^2}{D^2}+4\alpha^2}\bigg)$ --- the signs of $c$ and $\phi$ have been chosen so that $\theta \geq 0$ for $d \geq 1$. In the Eq. (\ref{eq:tilde_y_soln}), the unknown  constants $A$ and $B$ may be obtained from the relevant boundary and matching conditions, as we would show in the next section. From the exact $\tilde{F}(R,s)$, the exact $\langle T_{r} \rangle$ may be obtained through Eq.(\ref{eq:general_MFPT}), and studied as a function of the parameters of the model to understand its transitions. But the exact points of continuous  transition and tri-criticality may be obtained directly from the relationships among the moments (see Eqs (\ref{Conti_Trans}) and (\ref{Trans_TCP})).  In principle the moments may be obtained from $ \tilde{F}(R,s)$ as derivatives with respect to $s$, but the derivatives of $_1F_1$ and $U$ in Eq. (\ref{eq:tilde_y_soln}) do not lead to simple special functions. Hence, a better way to proceed is to solve for the moments directly from the differential equations they satisfy.  

%\par\texttt{Solution of moments for linear potential:} 
The $n^{th}$ moment is defined as $\langle T^n \rangle=\int^{\infty}_{0}t^{n}\bigg((-\frac{\partial Q}{\partial t})dt\bigg)=n\int^{\infty}_{0}t^{n-1}Qdt$. Multiplying  Eq. (\ref{eq_BFPE_1}) with $t^{n-1}$ and integrating over $[0,\infty)$ leads to the following recurrence equation: 
\be
D\frac{d^2\langle T^n \rangle}{dR^2}+\bigg(\frac{D(d-1)}{R}-k\bigg)\frac{d\langle T^n \rangle}{dR}=-n\langle T^{n-1} \rangle
\label{eq:Moments_eqn}
\ee
Thus if $\langle T^{n-1} \rangle$ is known, the moment $\langle T^{n} \rangle$ is obtained using Eq. (\ref{eq:Moments_eqn}). Note that zeroth order moment $\langle T^{0}\rangle =1$. The transformation $\xi^n=\frac{d\langle T^n\rangle}{dR}$ leads to first order linear differential equation of the form:
\be
\frac{d\xi^n}{dR}+\bigg(\frac{(d-1)}{R}-\frac{k}{D}\bigg)\xi^n=-\frac{n}{D}\langle T^{n-1}\rangle, 
\label{eq:Moments_eqn2}
\ee
which leads to the solution of $\xi^n$ 
\be
\xi^n(R)=R^{1-d} e^{(k/D) R}\bigg(A_n- \frac{n}{D}\int \langle T^{n-1}\rangle R^{'d-1} e^{-(k/D) R^{'}} dR^{'}\bigg)
\label{eq:Moments_eqn22}
\ee
with unknown constant $A_n$. Integrating $\xi^n(R)$ we get 
\be
\langle T^n \rangle=\int \xi^n(R') dR' +B_n
\label{eq:sol_moment}
\ee
where $B_n$ is another unknown constant. We use the above equations in Mathematica to solve for the relevant moments, and use them to find the transition points.

\section{\label{sec:sec4}Results for the model with simple drift under confinement}
For the model potential in Eq. (\ref{lin_pot}), we may obtain $\tilde{F}(R,s)$ by fixing the constants $A$ and $B$ in Eq. (\ref{eq:tilde_y_soln}), using the boundary conditions that $\tilde{F} = 1$ at $R=a$ and $R=R_m$. The exact expression is:

\hspace*{-2.0cm}\vbox{
\begin{eqnarray}
\tilde{F}(R,s)=&\frac{e^{\phi  \left(-a-R_m+R\right)}\big(\, _1F_1(\theta ;\gamma ;c R) (e^{a \phi } U(\theta,\gamma ,a c)-e^{\phi  R_m} U\left(\theta,\gamma ,c R_m\right))\big)}{U(\theta ,\gamma ,a c) \, _1F_1\left(\theta ;\gamma ;c R_m\right)-\, _1F_1(\theta;\gamma ;a c) U\left(\theta ,\gamma ,c R_m\right)}\nonumber\\
  &\bigg[1-\frac{U(\theta ,\gamma ,c R) \left(e^{a \phi } \, _1F_1(\theta ;\gamma ;a c)-e^{\phi  R_m} \, _1F_1\left(\theta ;\gamma ;c R_m\right)\right)}{\, _1F_1(\theta ;\gamma ;c R) (e^{a \phi } U(\theta ,\gamma ,a c)-e^{\phi  R_m} U\left(\theta ,\gamma ,c R_m\right))}\bigg]
  \end{eqnarray}
}

Replacing $R\to R_0$ and $s\to r$,  the above gives  $\tilde{F}(R_0,r)$ which when substituted in  Eq.(\ref{eq:general_MFPT}) gives the exact expression of $\langle T_r \rangle$. In particular for the study of the discontinuous transitions, this result is necessary. We plot the MFPT as a function of the resetting rate $r$ (for example see Fig.(\ref{fig:Model3_MFPT_vs_r}) in \ref{App_sec4}) and obtain the point and magnitude of the discontinuous jump in the ORR.

As we know $\langle T_r \rangle$ exactly as a function of $r$ and other parameters, we may plot it and obtain the global minima (ORR) and obtain the ORR vanishing transitions. This method is particularly suitable to locate the discontinuous transitions --- for an example, see Fig.(\ref{fig:Model3_MFPT_vs_r}) in \ref{App_sec4}. For the continuous transition and tri-critical points one may alternatively use the exact relations on moments  $\langle T^n \rangle$ (see Eq. (\ref{Conti_Trans}) and (\ref{Trans_TCP})) --- an example of this method is shown in Fig. (\ref{fig:2nd_1d_dim_linear_pot}) and (\ref{fig:TCP_d_dim_pw_linear_ordr}),  in \ref{App_sec4}. 
%To locate the points of continuous transition and tri-criticality  we solve Eq. (\ref{eq:sol_moment}) for $n=1$, $2$ and $3$. 
The associated constants $A_n$ and $B_n$ (in Eq. (\ref{eq:Moments_eqn22}) and (\ref{eq:sol_moment})) may be obtained by setting $\langle T^n \rangle=0$ at $R=a$ and $R=R_m$. The solutions are cumbersome expressions obtained using Mathematica and cannot be shown explicitly here. The resulting transitions will be discussed here. 
%The resulting continuous transitions, tri-critical points, and also the discontinuous transitions (using $\langle T_r \rangle$) will be discussed here.  
The relevant parameter space is formed by the potential strength $k$ and the initial/reset location $R_0$. In the following we show the transition lines in that parameter space for two sets of data --- one in which the dimension $d$ of the system varies at a fixed $\langle T_{on} \rangle$, and in another where $\langle T_{on}\rangle$ varies at a fixed dimension.

\subsection{Transitions for different $d$ at a fixed $\langle T_{on} \rangle$} 
\begin{figure}[hb!]
 \begin{subfigure}{0.325\textwidth}
  \includegraphics[width = 1\textwidth,height=0.16\textheight]{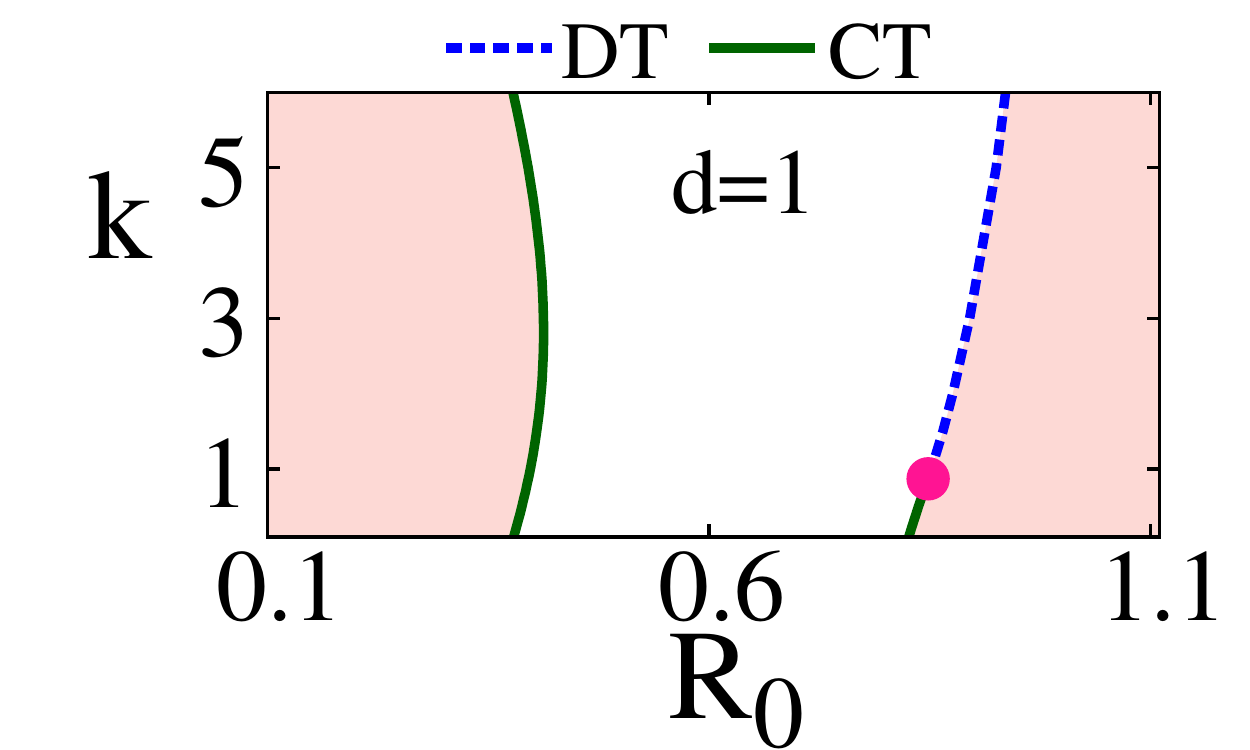}
  \caption[]{} 
  \label{fig:linear_pot_d_1_Ton_0}
 \end{subfigure}
 \begin{subfigure}{0.325\textwidth}
 \includegraphics[width = 1\textwidth,height=0.16\textheight]{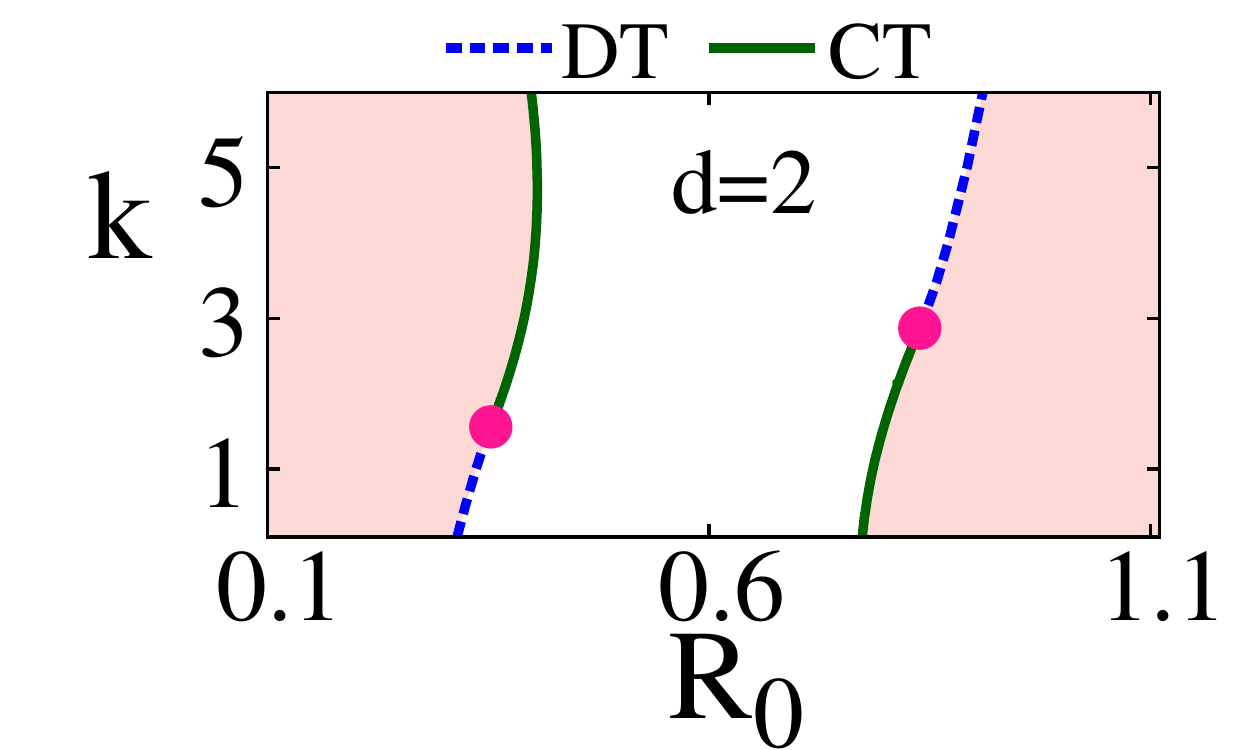}
  \caption[]{} 
  \label{fig:linear_pot_d_2_Ton_0}
\end{subfigure}
 \begin{subfigure}{0.325\textwidth}
   \includegraphics[width = 1\textwidth,height=0.16\textheight]{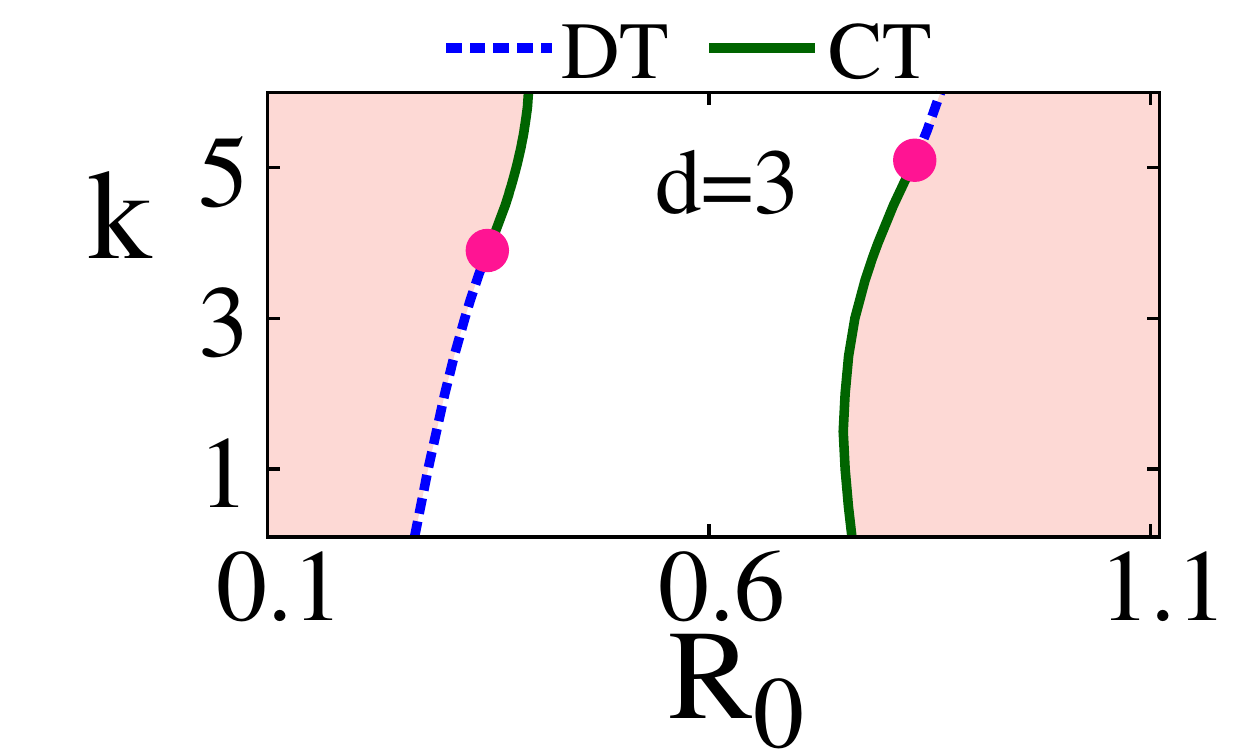}
  \caption[] {} 
  \label{fig:linear_pot_d_3_Ton_0}
 \end{subfigure}
 \caption{We show the ORR transition lines in the $k-R_0$ plane for different $d$ at a fixed $\langle T_{on}\rangle=0$. The shaded regions correspond to $r^*>0$. (a) For $d=1$ the tri-critical point (denoted by filled pink circle) is at $k^{2t}=0.88$, $R_{2c}=0.848$. In (b) a pair of tri-critical points are at $k^{1t}=1.50$, $R_{1c}=0.351$ and $k^{2t}=2.88$, $R_{2c}=0.839$ for $d=2$. In (c), for $d=3$, the tri-critical points move to ($k^{1t}=3.89$, $R_{1c}=0.349$ and $k^{2t}=5.03$, $R_{2c}=0.830$). We choose the parameters $D=1$, $a=0.1$, and $R_m=1.1$. Here DT and CT denote discontinuous and continuous transitions respectively.}
 \label{fig:linear_pot_w_d_Ton_0}
\end{figure}

We show the dynamical transitions of ORR in the $k-R_0$ plane for different $d$ by considering $\langle T_{on}\rangle=0$ in Fig. (\ref{fig:linear_pot_w_d_Ton_0}). In Fig. (\ref{fig:linear_pot_d_1_Ton_0}) we show the transitions for $d=1$ --- the results are known from \cite{PRR_Pal_Parsad_Landau_2019}.
%%%%%%%%%%%%%%%%%%%%%%%%
%%%%%%%%%%%%%%%%%%%%%%%%
%%%%%%%%%%%%%%%%%%%%%%%%
%\iffalse
%\fi
%%%%%%%%%%%%%%%%%%%%%%%%
%%%%%%%%%%%%%%%%%%%%%%%%
%%%%%%%%%%%%%%%%%%%%%%%%
Near the left boundary (for which the drift helps the motion), there is a continuous transition line throughout. While near the right boundary (for which the drift opposes the motion), the continuous line at small $k$ gives way to  a discontinuous line at large $k$ with a tri-critical point at $k^{2t}= 0.876$ and $R_{2c}=0.848$.

For $d=2$ (Fig. \ref{fig:linear_pot_d_2_Ton_0}),  there is the emergence of an additional tri-critical point ($k = k^{1t}$) near the inner absorbing boundary. For $k>k^{1t}$, the transitions are   continuous while $k<k^{1t}$ they are discontinuous. Note that the scenario for $k=0$ case was studied in \cite{PRE_Huang_higher_d_reset_2_ab_2022}.  For the outer boundary the second tri-critical point $k^{2t}$ remains, but is pushed up to a larger value of $k$ than in $d=1$. For $d=3$ (see Fig. \ref{fig:linear_pot_d_3_Ton_0}), both the tri-critical points shift upwards to the high $k$ direction, thus expanding the discontinuous transition line near the inner boundary and shrinking the corresponding discontinuous line near the outer boundary. 

\subsection{Transitions for different $\langle T_{on} \rangle$ at a fixed $d$:} 
\begin{figure}[hb!]
  \begin{subfigure}{0.325\textwidth}
    \includegraphics[width = 1\textwidth,height=0.15\textheight]{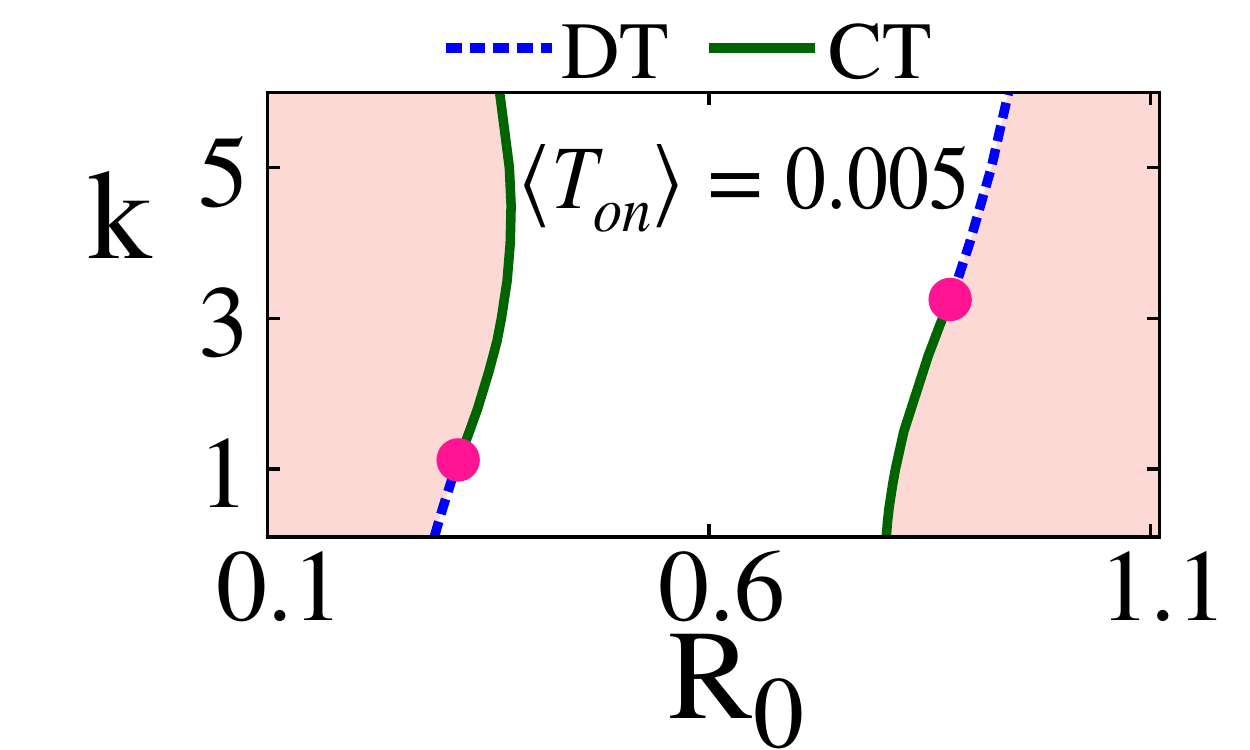}
    \caption[]{} 
    \label{fig:linear_pot_d_2_Ton_0.0}
   \end{subfigure}
  \begin{subfigure}{0.325\textwidth}
    \includegraphics[width = 1\textwidth,height=0.15\textheight]{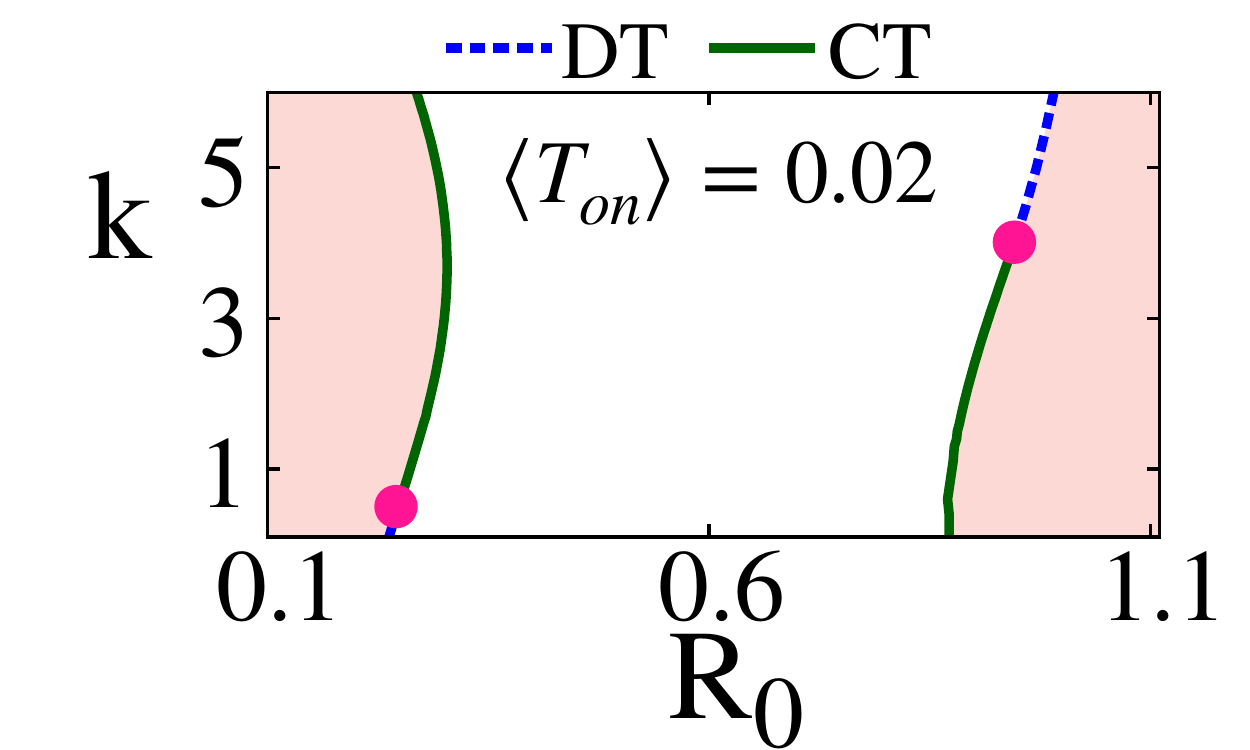}
    \caption[]{} 
    \label{fig:linear_pot_d_2_Ton_0.02}
  \end{subfigure}
   \begin{subfigure}{0.325\textwidth}
    \includegraphics[width = 1\textwidth,height=0.15\textheight]{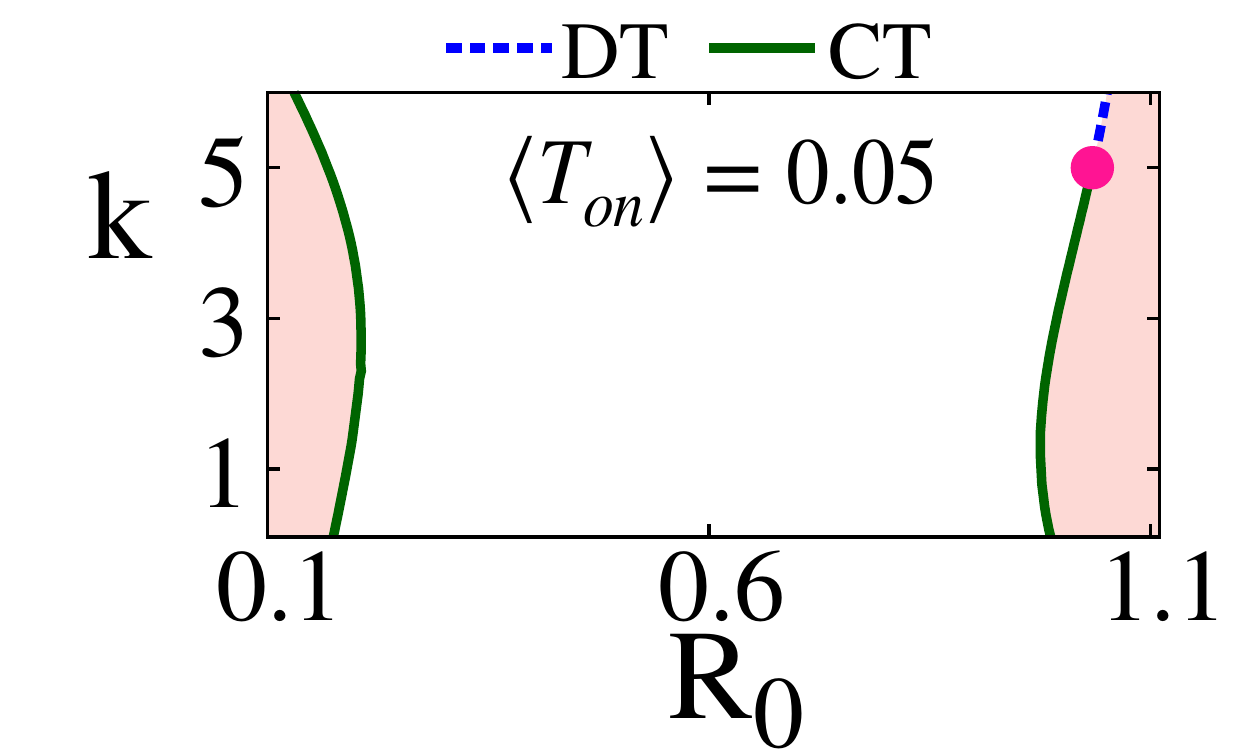}
   \caption[] {} 
    \label{fig:linear_pot_d_2_Ton_0.05}
  \end{subfigure}
   \caption{The ORR transition lines in the $k-R_0$ plane are shown for different $\langle T_{on} \rangle$ at a fixed $d=2$. The shaded regions are indicative of resetting beneficial zones (i.e. $r^*>0$). (a) For small $\langle T_{on} \rangle=0.005$, a pair of tri-critical points are located at $k^{1t}=1.13$, $R_{1c}=0.317$ and $k^{2t}=3.26$, $R_{2c}=0.873$. (b) For $\langle T_{on} \rangle =0.02$ they are at ($k^{1t}=0.43$, $R_{1c}=0.246$ and $k^{2t}=4.01$, $R_{2c}=0.946$). In (c) for $\langle T_{on} \rangle =0.05$, the left branch has no tri-critical point for any $k>0$ while the right one has a tri-critical point at $k^{2t}=5.03$, $R_{2c}=1.034$.  We choose $D=1$, $a=0.1$, and $R_m=1.1$ for these plots. Here DT and CT denote discontinuous and continuous transitions respectively.}
    \label{fig:linear_pot_d_2__w_Ton}
\end{figure}

We show the dynamical transitions of ORR in the $k-R_0$ plane for  different   $\langle T_{on}\rangle$ at fixed $d=2$ in  Fig. (\ref{fig:linear_pot_d_2__w_Ton}).  Overall, the regions of $r_* \neq 0$ reduce in the parameter space (compare the shaded regions in Fig. (\ref{fig:linear_pot_d_2__w_Ton}) to Fig. (\ref{fig:linear_pot_w_d_Ton_0})). Thus non-zero $\langle T_{on}\rangle$ helps the potential offset the benefit of resetting much more easily. 

%%%%%%%%%%%%%%%%%%
%%%%%%%%%%%%%%%%%%
%%%%%%%%%%%%%%%%%%
%\iffalse
%\fi
%%%%%%%%%%%%%%%%%%
%%%%%%%%%%%%%%%%%%
%%%%%%%%%%%%%%%%%%

At small value of $\langle T_{on}\rangle = 0.005$  (see Fig. (\ref{fig:linear_pot_d_2_Ton_0.0})), the scenario is very similar to Fig. (\ref{fig:linear_pot_d_2_Ton_0}). There are two tri-critical points ($k^{1t}$ and $k^{2t}$) and discontinuous transition lines exist near both the inner and outer absorbing boundaries. As $\langle T_{on} \rangle$ increases, for the left branch, the segment of discontinuous transition line shrinks in length with the $k^{1t}$ shifting to lower values of $k$ -- see the Fig. (\ref{fig:linear_pot_d_2_Ton_0.02}) for $\langle T_{on} \rangle = 0.02$. Beyond a certain  value of delay time  $\langle T_{on} \rangle^{*} = 0.034$ there is no tri-critical point at any $k >0$ near the inner boundary --- see the continuous transition line in the left side of Fig. (\ref{fig:linear_pot_d_2_Ton_0.05}) for $\langle T_{on} \rangle = 0.05$.

If one compares Figs. (\ref{fig:linear_pot_d_2__w_Ton}) and (\ref{fig:linear_pot_w_d_Ton_0}), one would see that for the left branch of ORR transitions near the inner boundary, the effect of increasing $d$ is just the opposite to that of increasing $\langle T_{on} \rangle$ -- the Fig. (\ref{fig:linear_pot_d_2_Ton_0.05}) resembles the Fig. (\ref{fig:linear_pot_d_1_Ton_0}) qualitatively.  On the other hand the right branch of transitions seem to evolve similarly in the two cases of rising $d$ and $\langle T_{on} \rangle$, as $k^{2t}$ moves to higher values of $k$.

\section{\label{sec:sec5}Results for the problem with a barrier potential in confined space} 
In this part we study the transition in ORR for a piece-wise linear potential with a barrier (Eq. (\ref{Tent_pot})).  The solution of $\tilde{F}(R,s)$ in Eq. (\ref{eq:tilde_y_soln}) for a linear potential, may be incorporated to obtain the solutions for the piece-wise segments separately. Let  $\tilde{F} = \tilde{F}_{1}$ in the region  $R\in [a,R_t)$ and $\tilde{F} = \tilde{F}_2$ in the region $R\in (R_t,R_m]$ --- there are four unknown constants, a pair for $\tilde{F}_1$ and a pair for $\tilde{F}_2$ according to Eq. (\ref{eq:tilde_y_soln})) (see appendix Eqn. (\ref{eq:tilde_y_soln_l_1}) and \ref{eq:tilde_y_soln_r_1}). These may be fixed by using two absorbing boundary conditions, namely $\tilde{F}_1=1$ at $R=a$ and $\tilde{F}_2=1$ at $R=R_m$, and two matching conditions at $R=R_t$, namely $\tilde{F}_1=\tilde{F}_2$ and $\tilde{F}^{'}_1=\tilde{F}^{'}_2$ (see appendix Eq. (\ref{App_eq:Lap_FPT_Match})). We have used Mathematica to handle these cumbersome expressions, and obtain $\tilde{F}_1$ and $\tilde{F}_2$ fully, which in turn gives the MFPT ($\langle T_r \rangle_1$ and $\langle T_r \rangle_2$) to reach either of the targets starting from any $R$, with resetting and $\langle T_{on} \rangle$ in any $d$, using Eq.(\ref{eq:general_MFPT}).  This is particularly helpful to locate the discontinuous transitions.  On the other hand the continuous and tri-critical points may be obtained from the moment equalities  in Eqs (\ref{Conti_Trans}) and (\ref{Trans_TCP}).  Let the moments be $\langle T^{n} \rangle_1$ and $\langle T^{n} \rangle_2$ for $R\leq R_t$ and $R\geq R_t$ respectively. The $n^{th}$ moment satisfy the following four conditions $\langle T^{n} \rangle_1=0$ at $R=a$, $\langle T^{n} \rangle_2=0$ at $R=R_m$ and $\langle T^{n} \rangle_1=\langle T^{n}\rangle_2$ and $\langle T \rangle^{'}_1=\langle T \rangle^{'}_2$ (see appendix Eq. (\ref{App_eq:MFPT_Match})). We needed only moments for $n=1,2,$ and $3$ for our calculations of the exact critical points.

\subsection{Transitions for different $d$ at a fixed $\langle T_{on} \rangle$:} 
\begin{figure}[ht!]
  \begin{subfigure}{0.325\textwidth}
    \includegraphics[width = 1\textwidth,height=0.15\textheight]{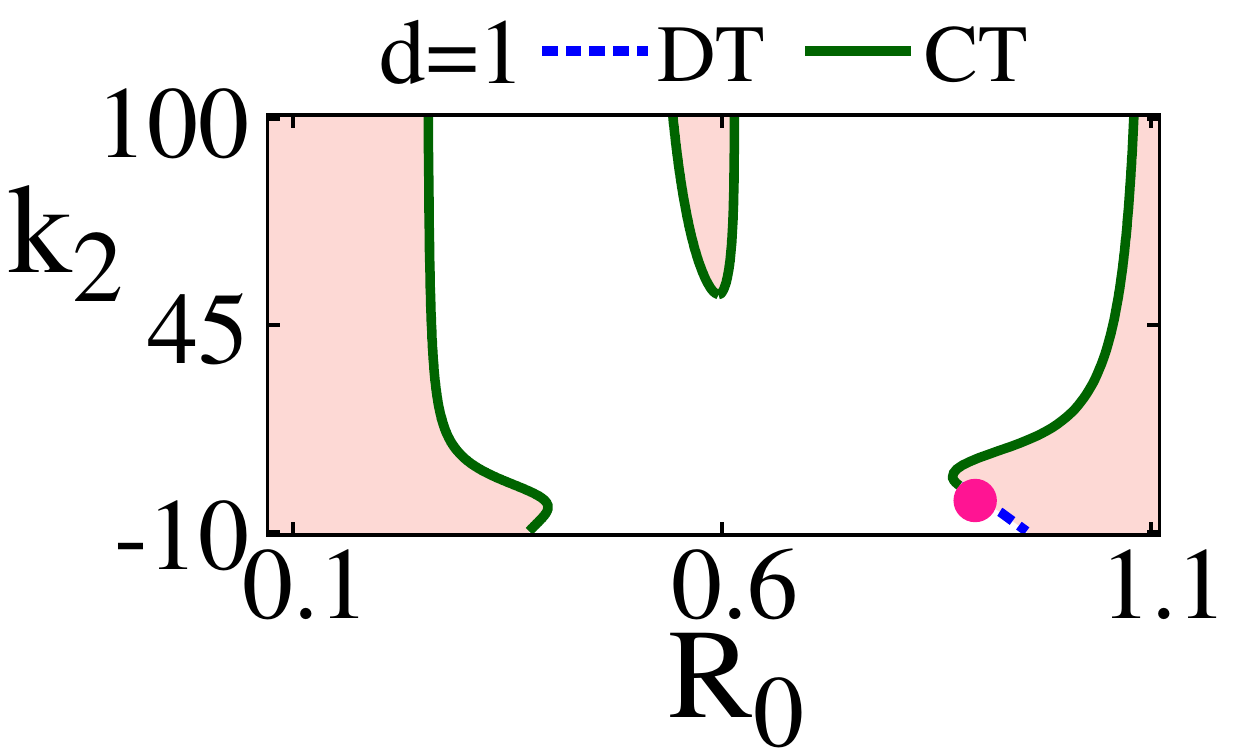}
    \caption[]{} 
    \label{fig:Tent_pot_d_1_Ton_0}
   \end{subfigure}
  \begin{subfigure}{0.325\textwidth}
    \includegraphics[width = 1\textwidth,height=0.15\textheight]{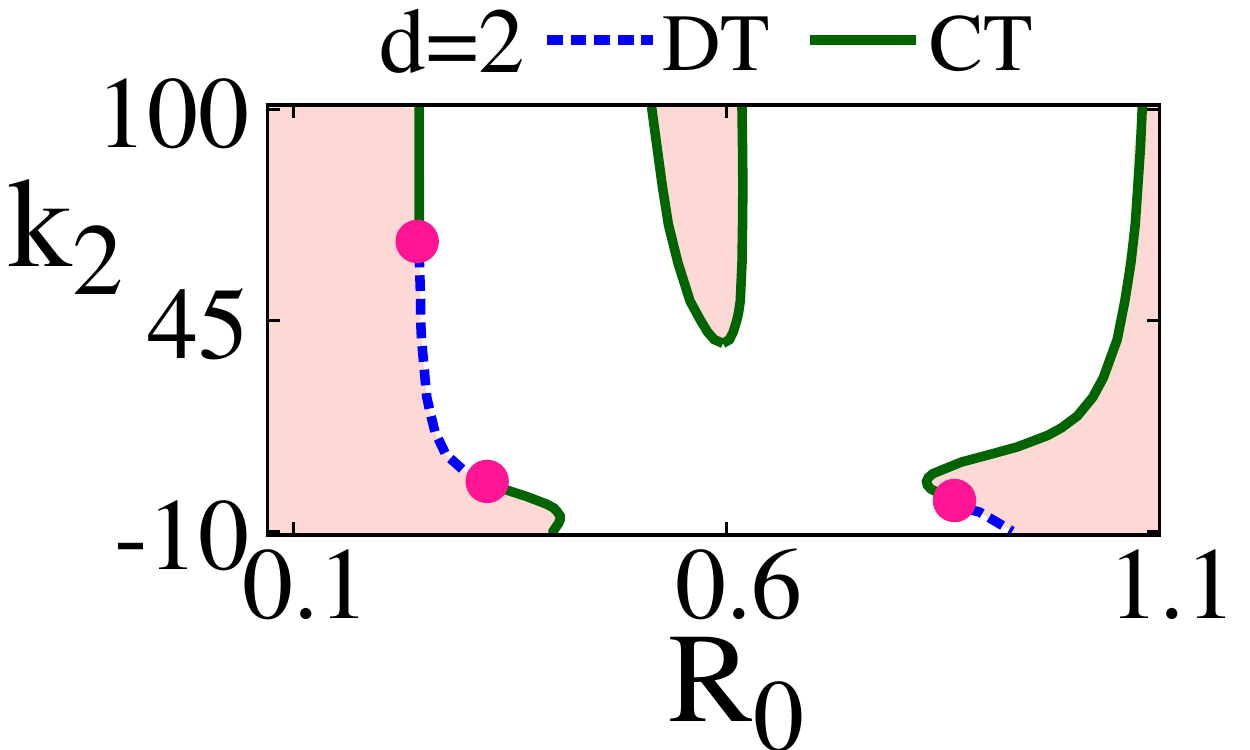}
    \caption[]{} 
    \label{fig:Tent_pot_d_2_Ton_0}
  \end{subfigure}
   \begin{subfigure}{0.325\textwidth}
    \includegraphics[width = 1\textwidth,height=0.15\textheight]{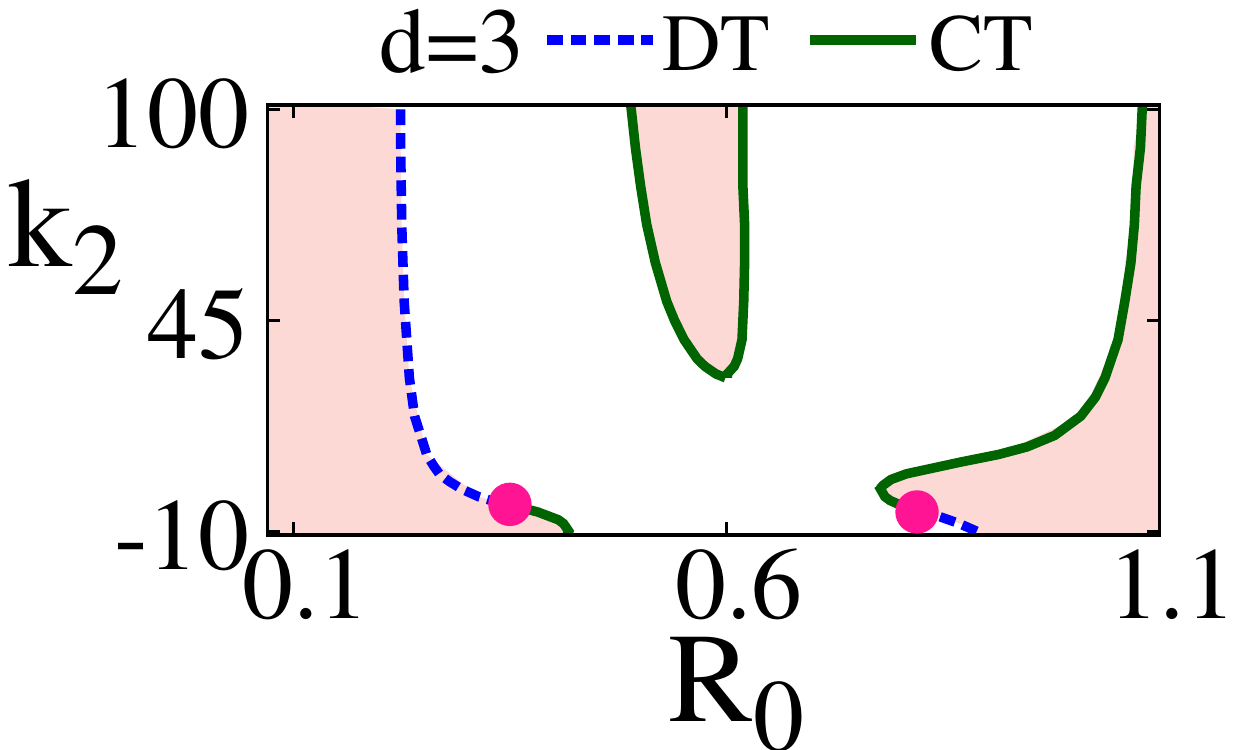}
   \caption[] {} 
    \label{fig:Tent_pot_d_3_Ton_0}
  \end{subfigure}
   \caption{The ORR transition lines are shown in the $k_2-R_0$ plane for different $d$ at a fixed $\langle T_{on} \rangle =0.0$ with $k_1=5$. In (a) one tri-critical point is at ($k^{2t}_2=2.11$, $R_{2c}=0.877$) for $d=1$. (b) For $d=2$ the tri-critical points are at ($k^{1t}_2=2.97$, $R_{1c}=0.318$ and $k^{3t}_2=65.01$, $R_{3c}=0.245$) on the left branch and at $k^{2t}_2=-1.07$, $R_{2c}=0.855$ on the right. (c) For $d=3$ a tri-critical point is at $k^{1t}_2=-2.7$, $R_{1c}=0.341$ and the other one on the left branch is at values of $k_2 > 100$ and hence not seen in the figure; the right one is at $k^{2t}_2=-5.08$, $R_{2c}=0.830$. The parameters used are mentioned in the text. Here DT and CT denote discontinuous and continuous transitions respectively and the shaded regions are with $r^*>0$.}
 \label{fig:Tent_pot_w_d_Ton_0}
\end{figure}

We show in Fig. (\ref{fig:Tent_pot_w_d_Ton_0})  the dynamical transitions in the $k_2-R_0$ plane for different dimensions $d$ at fixed $\langle T_{on} \rangle=0$ and $k_1$. For $d=1$ the transitions have been elaborately studied in \cite{PRE_Saeed_2022}. In Fig. (\ref{fig:Tent_pot_d_1_Ton_0}) we reproduce the case of $k_1=5$, $a=0.1$, $R_t = 0.6$ and $R_m=1.1$. The left branch near the boundary $R=a$ is a continuous transition line, while near the boundary $R=R_m$, there is a tri-critical point at $k^{2t}_2$. For $k_2<k^{2t}_2$ the transition is discontinuous while for $k_2>k^{2t}_2$ it is continuous. 
%%%%%%%%%%%%%%%%%%
%%%%%%%%%%%%%%%%%%
%%%%%%%%%%%%%%%%%%
%\iffalse
%\fi
%%%%%%%%%%%%%%%%%%
%%%%%%%%%%%%%%%%%%
%%%%%%%%%%%%%%%%%%
  For large positive value of $k_2$ there is an island of resetting beneficial region, hence at a fixed $k_2$ as a function of $R_0$ there are four critical points. 

For the same parameters $k_1, a, R_t, R_m$ mentioned above, we study the problem in higher dimensions. For $d=2$ (Fig. (\ref{fig:Tent_pot_d_2_Ton_0})), for the left branch, there is an emergence of a line segment of discontinuous transition in ORR  (shown in dashed line) with two tri-critical points at $k_2=k^{1t}_2$ and $k_2=k^{3t}_2$ marking the two endpoints. The right branch and the island at larger $k_2$ change slightly such that the area of $r_* \neq 0$ increases a bit in comparison to $d=1$. Increasing the  dimension further to $d=3$ we see in Fig. (\ref{fig:Tent_pot_d_3_Ton_0}), that the length of the discontinuous line segment increases such that the upper tri-critical point moves out of the range of the figure while the  lower one moves a bit below. The area of $r_* \neq 0$ bounded by the right branch and the central island increases further. 

Thus overall with rising dimensions, the regions of space where resetting is beneficial (i.e. $r_* > 0$) increases, and secondly, the tendency of discontinuous transition in ORR enhances, for resetting to locations near the left absorbing boundary.

\subsection{Transitions for different $\langle T_{on} \rangle$ at a fixed $d$:}
\begin{figure}[ht!]
  \begin{subfigure}{0.325\textwidth}
    \includegraphics[width = 1\textwidth,height=0.15\textheight]{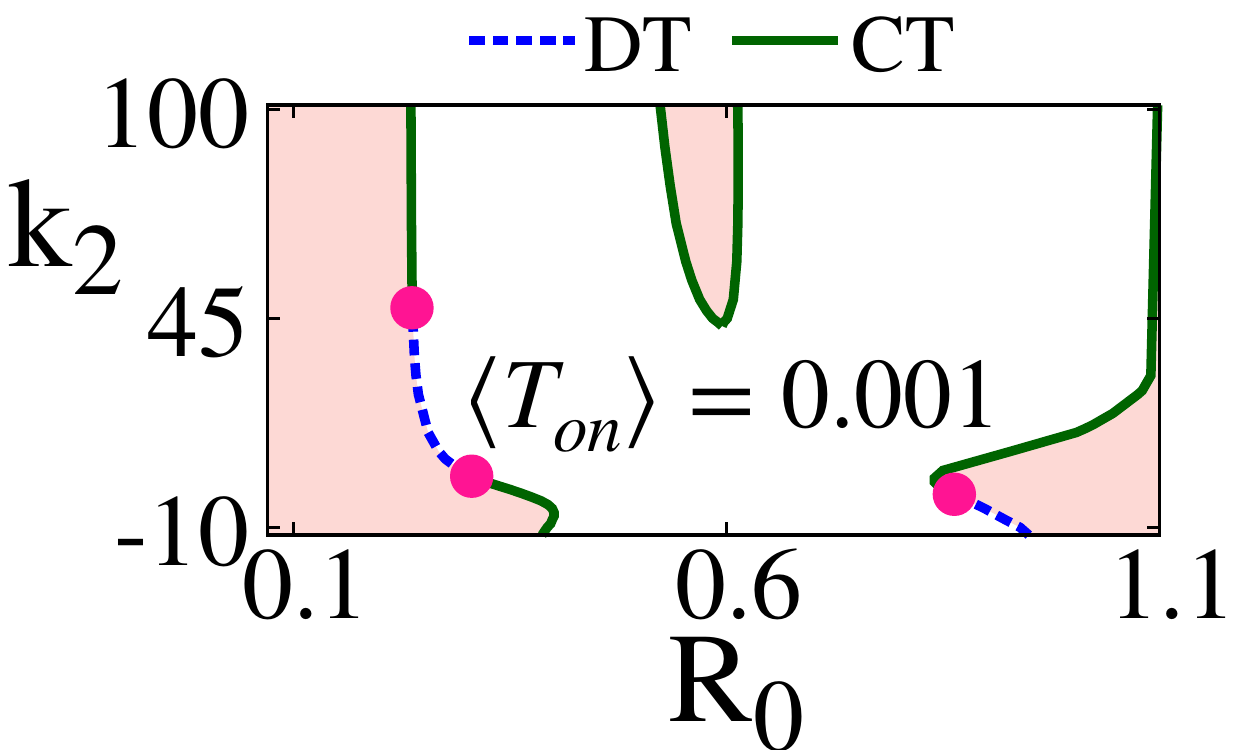}
    \caption[]{} 
    \label{fig:Tent_pot_d_2_Ton_0.0}
   \end{subfigure}
  \begin{subfigure}{0.325\textwidth}
    \includegraphics[width = 1\textwidth,height=0.15\textheight]{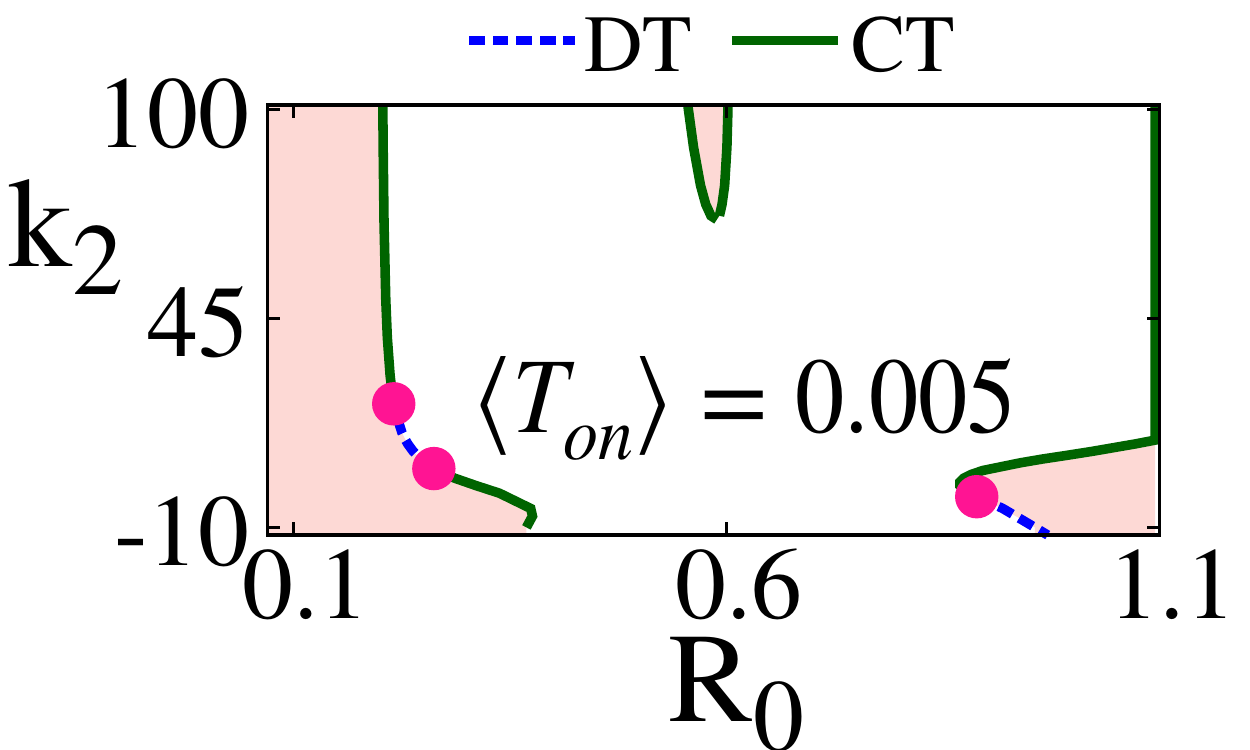}
    \caption[]{} 
    \label{fig:Tent_pot_d_2_Ton_0.005}
  \end{subfigure}
   \begin{subfigure}{0.325\textwidth}
    \includegraphics[width = 1\textwidth,height=0.15\textheight]{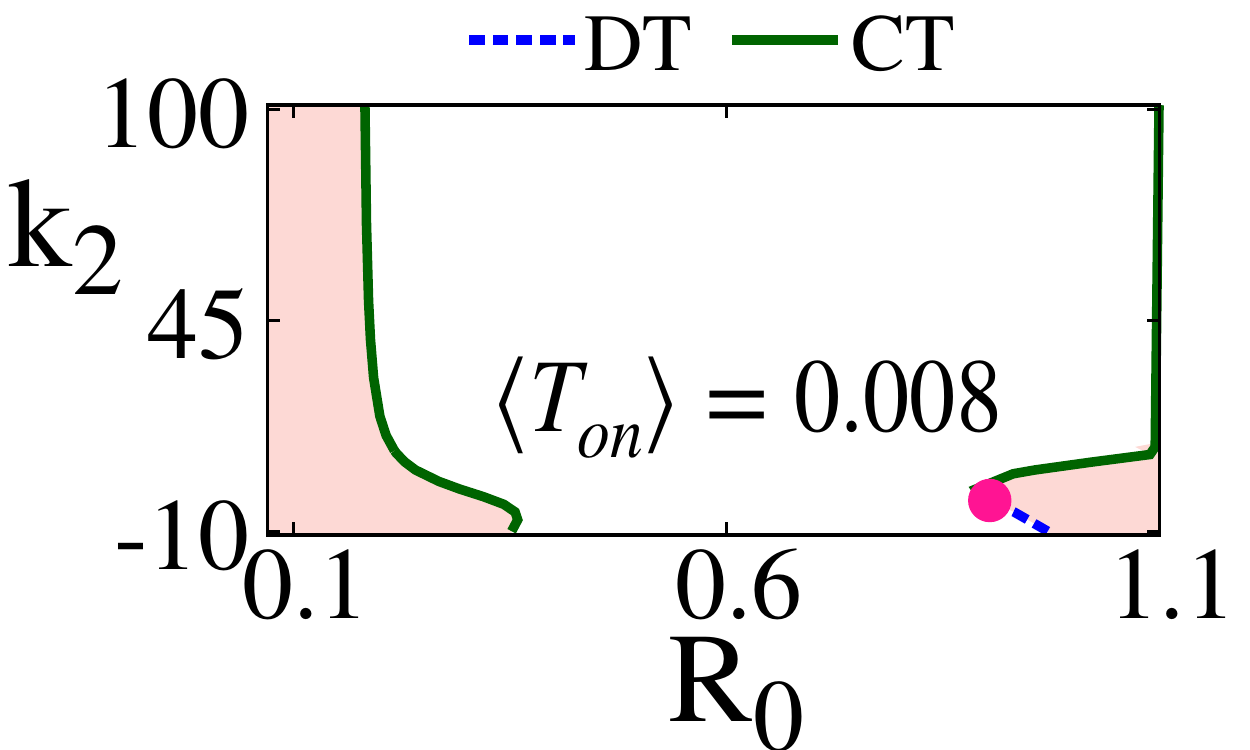}
   \caption[] {} 
    \label{fig:Tent_pot_d_2_Ton_0.008}
  \end{subfigure}
   \caption{We show the ORR transition lines in the $k_2-R_0$ plane for different $\langle T_{on} \rangle$ at a fixed $d=2$ with $k_1=5$. (a) For $\langle T_{on}\rangle=0.001$, on the left branch the tri-critical points are at $k^{1t}_2=3.44$, $R_{1c}=0.306$, and $k^{3t}_2=48.38$, $R_{3c}=0.237$, and on the right at $k^{2t}_2=-1.26$, $R_{2c}=0.863$. (b) For $\langle T_{on} \rangle=0.005$ tri-critical points are at $k^{1t}_2=5.61$, $R_{1c}=0.261$,  and $k^{3t}_2=22.49$, $R_{3c}=0.216$ and $k^{2t}_2=-1.88$, $R_{2c}=0.889$. In (c) for $\langle T_{on}\rangle=0.008$, a single tri-critical point is at $k^{2t}_2=-2.26$, $R_{2c}=0.906$. The parameters used are mentioned in the text. Here DT and CT denote discontinuous and continuous transitions respectively. The shaded regions are with $r^*>0$.}
    \label{fig:Tent_pot_d_w_Ton}
\end{figure}

In Fig. (\ref{fig:Tent_pot_d_w_Ton}) we show the transitions of the ORR in the $k_2-R_0$ plane by varying  $\langle T_{on} \rangle$ at a fixed $d=2$ and $k_1=5$. In Fig.(\ref{fig:Tent_pot_d_2_Ton_0.0}) we show the case for $\langle T_{on} \rangle=0.001$ which is very close to the  Fig.(\ref{fig:Tent_pot_d_2_Ton_0}). 
%%%%%%%%%%%%%%%%%%
%%%%%%%%%%%%%%%%%%
%%%%%%%%%%%%%%%%%%
%\iffalse
%\fi
%%%%%%%%%%%%%%%%%%
%%%%%%%%%%%%%%%%%%
%%%%%%%%%%%%%%%%%%
  There are three tri-critical points (two near the inner boundary $a$ and one near the outer boundary $R_m$). The continuous and discontinuous lines are shown in solid and dashed lines respectively. 

Increasing $\langle T_{on}\rangle$ by a small amount to $\langle T_{on}\rangle =0.005$ (see Fig.(\ref{fig:Tent_pot_d_2_Ton_0.005})), in the left branch, the two tri-critical points approach each other.  Moreover the areas of the island region and the one bounded by the critical branch near $R_m$ shrink considerably. For $\langle T_{on} \rangle=0.008$ (in Fig.(\ref{fig:Tent_pot_d_2_Ton_0.008})), there is only a line of continuous transition on the left side as the two tri-critical points have merged. Moreover, the island region has moved to higher $k_2$, out of  the frame of the figure. 

Thus overall for the barrier potential, the effect of rising $\langle T_{on}\rangle$ seems to be exactly the opposite of increasing $d$. We see that the space where resetting is beneficial (i.e. $r_* > 0$) decreases, and the tendency of discontinuous transition in ORR diminishes with increasing $\langle T_{on}\rangle$. The tri-critical points on the left branch vanish beyond a threshold value of $\langle T_{on}\rangle$. 

\section{\label{sec:sec6}Discussion}
In this article we analytically studied the problem of first passage of a diffusing particle  under stochastic reset with delayed restart, under a potential, in arbitrary dimensions.  The geometry is a confined space between two concentric spheres, both of which serve as possible targets. We studied the problem for two types of potentials, one which provides an ordinary drift towards the inner radius $a$, while another which offers a barrier in between the radii $a$ and $R_m$. In both the cases, the effect of delayed restart ($\langle T_{on}\rangle \neq 0$) is to enhance the MFPT (see Eq. (\ref{eq:general_MFPT})), and suppress the advantage of resetting which make the areas  of non-zero ORR decrease in Figs. (\ref{fig:linear_pot_d_2__w_Ton}) and (\ref{fig:Tent_pot_d_w_Ton}). Interestingly, while the possibility of discontinuous transition near the inner boundary increases in higher dimensions for both the potentials (Figs. (\ref{fig:linear_pot_w_d_Ton_0}) and (\ref{fig:Tent_pot_w_d_Ton_0})),  the effect is the reverse  with rising  $\langle T_{on}\rangle$ (Figs. (\ref{fig:linear_pot_d_2__w_Ton}) and (\ref{fig:Tent_pot_d_w_Ton})). 

To understand the effect of $\langle T_{on}\rangle$ on the transitions we may look at Eqs. (\ref{Conti_Trans}) and (\ref{Trans_TCP}).  Note that for continuous transitions in the absence of $\langle T_{on}\rangle$, the variance   
$\sigma^2$ has to match $\langle T\rangle^2$. But for $\langle T_{on}\rangle \neq 0$, there is an additional term 
$2\langle T\rangle \langle T_{on}\rangle$ on the right side -- this makes the equality harder to occur, and hence we 
see the critical lines pushed back leaving more regions where resetting is non-beneficial (i.e. $r_* = 0$). 
For the tri-critical point, the balance is between the third moment and a product of second order fluctuations and the mean -- see  Eq. (\ref{Trans_TCP}). The equality for occurrence of tri-criticality has an additional positive term  $6\langle T\rangle \langle T_{on}\rangle^2$ on the right, for $\langle T_{on}\rangle \neq 0$. Thus, this equality will be harder to satisfy, and hence the continuous lines would not terminate at critical points easily,  such that 
the discontinuous transitions would become rarer. This is what we see in Figure  (\ref{fig:Tent_pot_d_w_Ton}).

Next we try to understand physically the role of the dimensions. The discontinuous transition with an abrupt jump in ORR is a manifestation of the urgent necessity of the system to take advantage of resetting to attain first passage. Note that for the drift potential in 1-d, the right target where, diffusive trajectories are less likely to reach in comparison to the left target, has the discontinuous transition near it. The uphill to the right with rising slope $k$ makes transport hard, and in its vicinity demands a sudden advantage through resetting.  As the dimension increases, the space near the outer boundary at $R_m$ is enhanced and its absorption likelihood increases by diffusion itself, in a way reducing the challenge of the uphill potential towards that boundary.  Hence understandably, the discontinuous transition line shortens near the outer boundary and increases near the inner boundary (see   Fig. (\ref{fig:linear_pot_w_d_Ton_0})).  Similarly we may understand the behavior for the tent potential in Fig.  (\ref{fig:Tent_pot_w_d_Ton_0}). Note that in this case $k_1$ is fixed, so with increasing $k_2$, the inner boundary at $a$ has to be accessed by crossing a steeper barrier. This task becomes increasing more challenging with rising dimensions as space near $R_m$ is increasingly more than near $a$ to accommodate diffusive excursions. Thus discontinuous transitions become an urgent necessity, and we see the length of the discontinuous line increases near the inner boundary in  Fig.  (\ref{fig:Tent_pot_w_d_Ton_0}).  

Recent experiments with optical tweezers and traps have verified several theoretical predictions of resetting \cite{J_Chem_letter_Reuveni_experimental-reset_2020,PRR_Experiment_evidence_Majumdar_2020}. In future if multiple such traps can be used to create potential minima separated by barriers, and first passage under resetting in such a system is studied, some of the interesting predictions on transitions of ORR that we make here and our earlier work \cite{PRE_Saeed_2022} may be verified.

\section*{Acknowledgments}
D.D. would like to acknowledge Science and Engineering Research Board (SERB) India (Grant No.MTR/2019/000341) for financial support. S.A. thanks IIT Bombay for the Institute Ph.D. fellowship.
\appendix

\section{\label{App_sec11}}
Consider a point particle having diffusivity $D$ in a $d-$dimensional space, under an attractive potential $V(\vec{R})$. The first passage problem may be developed through the standard formalism of backward Fokker-Planck equation \cite{gard_book_2004} for the survival probability $Q(\vec{R},t)$ for the particle to survive up to time $t$ starting from $\vec{R}$:
\begin{equation}
\frac{\partial Q(\vec{R},t)}{\partial t}= D\nabla^{2} Q(\vec{R},t)-\vec{\nabla}V(\vec{R}).\vec{\nabla}Q(\vec{R},t)
\label{eq:gen_back_FK_eqn_d}
\end{equation}
For radially symmetric potentials $V(\vec{R})=V(|\vec{R}|)$ and radially symmetric boundary conditions $Q(|\vec{a}|,t)=0$, $Q(|\vec{R_m}|,t)=0$ and initial condition $Q(|\vec{R}|,0)=1$, we need to only consider the radial part of the Laplacian and gradient operators. This simplifies the Eq.(\ref{eq:gen_back_FK_eqn_d}) to lead to:
\begin{equation}
\hspace{-0.4cm}\frac{\partial Q(R,t)}{\partial t}=D\frac{\partial^2 Q(R,t)}{\partial R^2}+\bigg(\frac{D(d-1)}{R}-V^{'}(R)\bigg) \frac{\partial Q(R,t)}{\partial R}.
\label{App:eq_BFPE_1}
\end{equation}
where $V^{'}(R)=\frac{dV(R)}{dR}$. In the manuscript in Eq.(8) we have assumed $V(R)$ equal to Eq.(\ref{lin_pot}) such that $V^{'}(R)=k$.

\section{\label{App_sec1}}
Here we discuss the matching conditions for the piece-wise solutions of the survival probabilities $Q_1(R,t)$ and $Q_2(R,t)$ over the two intervals $[0,R_t)$ and $(R_t,R_m]$ across $R=R_t$. This is used for the model with a barrier potential.

For the piece-wise linear potential (Eq. (\ref{Tent_pot})), the backward equation (Eq.(\ref{eq_BFPE_1})) may be written as follows:
\begin{equation}
\hspace{-2.8cm}\frac{\partial Q(R,t)}{\partial t}=D\frac{\partial^2 Q(R,t)}{\partial R^2}+\bigg(\frac{D(d-1)}{R}-(k_1-(k_2+k_1)\Theta(R-R_t))\bigg) \frac{\partial Q(R,t)}{\partial R}
\label{App_eq_BFPE_3}
\end{equation}
where, $\Theta$ represents the Heaviside step function \cite{Arfken_7ed_book}.

Integrating both sides of Eq.(\ref{App_eq_BFPE_3}) from $R=R_t-\epsilon$ to $R=R_t+\epsilon$, where $\epsilon$ is an infinitesimally small positive number, we have:
\hspace*{-2.0cm}\vbox{
\begin{eqnarray*}
  \nonumber\int_{R_t-\epsilon}^{R_t+\epsilon}dR &\frac{\partial Q(R,t)}{\partial t}=\int_{R_t-\epsilon}^{R_t+\epsilon}dRD\frac{\partial^2 Q(R,t)}{\partial R^2}\\
  & +\int_{R_t-\epsilon}^{R_t+\epsilon}dR\bigg(\frac{D(d-1)}{R}-(k_1-(k_2+k_1)\Theta(R-R_t))\bigg) \frac{\partial Q(R,t)}{\partial R}\textcolor{white}{---}
\label{App_eq_BFPE_4}
  \end{eqnarray*}
}

Since $Q(R,t)$ and $\frac{\partial Q(R,t)}{\partial t}$ are finite everywhere, it implies $\int_{R_t-\epsilon}^{R_t+\epsilon}Q(R,t)dR=2\epsilon Q(R_t,t)\rightarrow 0$ and $\int_{R_t-\epsilon}^{R_t+\epsilon}dR \frac{\partial Q(R,t)}{\partial t}=2\epsilon  \frac{\partial Q(R_t,t)}{\partial t}\to 0$. %Moreover the $\epsilon\to 0$ such that $R\to R_t$. 
Then Eq. (\ref{App_eq_BFPE_4}) reduces to
\hspace*{-3.0cm}\vbox{
\begin{eqnarray*}
-\int_{R_t-\epsilon}^{R_t+\epsilon}dRD\frac{\partial^2 Q(R,t)}{\partial R^2}
  &=\int_{R_t-\epsilon}^{R_t+\epsilon}dR\bigg(\frac{D(d-1)}{R_t}-(k_1-(k_2+k_1)\Theta(R-R_t))\bigg)  \frac{\partial Q(R,t)}{\partial R}\nonumber\\
&=\int_{R_t-\epsilon}^{R_t+\epsilon}dR\bigg(\frac{D(d-1)}{R_t}\bigg)\frac{\partial Q(R,t)}{\partial R}+k_1\int_{R_t-\epsilon}^{R_t}dR\frac{\partial Q(R,t)}{\partial R}\nonumber\\
&-k_2\int_{R_t}^{R_t+\epsilon}dR\frac{\partial Q(R,t)}{\partial R}
\label{App_eq_BFPE_5}
\end{eqnarray*}}

This gives

\hspace*{-3.0cm}\vbox{
\begin{eqnarray}
 - D(\frac{\partial Q(R,t)}{\partial R}\Bigr|_{R=R_t+\epsilon}-\frac{\partial Q(R,t)}{\partial R}\Bigr|_{R=R_t-\epsilon})=&\bigg(\frac{D(d-1)}{R_t}\bigg)(Q(R_t+\epsilon,t)-Q(R_t-\epsilon,t))\nonumber\\
  &+k_1(Q(R_t,t)-Q(R_t-\epsilon,t))\nonumber\\
  &-k_2(Q(R_t+\epsilon,t)-Q(R_t,t))
\end{eqnarray}}

Assuming $Q(R,t)$ is continuous through $R=R_t$, the right hand side of the above equation vanishes and we see that the first derivative  $\frac{\partial Q(R,t)}{\partial R}$ is also continuous. Thus we have
\begin{equation}
Q_1(R_t,t)=Q_2(R_t,t)\textcolor{white}{---}\text{and}\nonumber\\
Q^{'}_1(R_t,t)=Q^{'}_2(R_t,t)
\end{equation} 
In the Laplace space these matching conditions give:
\begin{equation*}
  q_1(R_t,s)=q_2(R_t,s)\textcolor{white}{----}\text{and}\textcolor{white}{---}\\  
  q^{'}_1(R_t,s)=q{'}_2(R_t,s)
    \label{App_eq:Lap_survive_Match}
\end{equation*}
Similarly the first passage distribution defined as $\tilde{F}(R,s)=1-sq(R,s)$ is
\be
  \tilde{F}_1(R_t,s)=\tilde{F}_2(R_t,s)\textcolor{white}{---}\text{and}\nonumber\\
  \tilde{F}^{'}_1(R_t,s)=\tilde{F}^{'}_2(R_t,s)
    \label{App_eq:Lap_FPT_Match}
\ee
Since the $n^{th}$ moment $\langle T_r^{n} \rangle =n(-1)^{n-1}\frac{\partial^{n-1} q(R_0,s)}{\partial s^{n-1}}\bigg|_{s\to0}$ the matching conditions at $R=R_t$ for the moments are:
\be
    \langle T_r^{n} \rangle_{1}=\langle T_r^{n} \rangle_{2}\textcolor{white}{----}\text{and}\textcolor{white}{---}\\
    \langle T_r^{n} \rangle^{'}_{1}=\langle T_r^{n} \rangle^{'}_{2}
    \label{App_eq:MFPT_Match}
    \ee

 \section{\label{App_sec2}}
 The solution of $\tilde{F}(R,s)$ for linear segment is
\begin{equation}
  \tilde{F}(R,s) =Ae^{\phi R}\, _1F_1(\theta ;\gamma ;c R)+Be^{\phi R}U(\theta,\gamma,cR)
  \label{eq:App_tilde_y_soln}
\end{equation}
where $c=\sqrt{\frac{k^2}{D^2}+4\alpha^2}$, $\gamma =d-1$, $\theta=-\frac{\phi(d-1)}{c}$ and $\phi=\frac{1}{2}\bigg(\frac{k}{D}-\sqrt{\frac{k^2}{D^2}+4\alpha^2}\bigg)$. For the piece-wise linear barrier  potential (Eq. (\ref{Tent_pot})) the Eq. (\ref{eq:App_tilde_y_soln}) in region $R\in[a,R_t)$ is
   \begin{equation}
  \tilde{F}_1(R,s) =A_1e^{\phi R}\, _1F_1(\theta ;\gamma ;c R)+B_1e^{\phi R}U(\theta,\gamma,cR)
  \label{eq:tilde_y_soln_l_1}
   \end{equation}
with $k = k_1$ and $A_1$, $B_1$ are constants. For $R\in(R_t,R_m]$, $\tilde{F}(R,s)=\tilde{F}_2(R,s)$ implies
  \begin{equation}
  \tilde{F}_2(R,s) =A_2e^{\phi R}\, _1F_1(\theta ;\gamma ;c R)+B_2e^{\phi R}U(\theta,\gamma,cR)
  \label{eq:tilde_y_soln_r_1}
  \end{equation}
  with $k =  -k_2$ and constants $A_2$ and $B_2$. These four constants ($A_1$, $B_1$, $A_2$, and $B_2$) may be obtained using two boundary conditions $\tilde{F}_1(R=a,s)=1$, $\tilde{F}_2(R=R_m,s)=1$, and the two matching conditions Eq. (\ref{App_eq:Lap_FPT_Match}) at $R=R_t$. We solve these unknowns using Mathemetica and substitute in Eqn. (\ref{eq:tilde_y_soln_l_1}) and (\ref{eq:tilde_y_soln_r_1}) to get the desired $\tilde{F}_1(R,s)$, $\tilde{F}_2(R,s)$.

  Similarly solution of $n^{th}$ moment without resetting on the both sides of $R_t$ may be obtained using Eq. (\ref{eq:sol_moment}).

\section{\label{App_sec4}}
\begin{figure}[ht!]
  \centering
  \includegraphics[width = 0.65\textwidth,height=0.3\textheight]{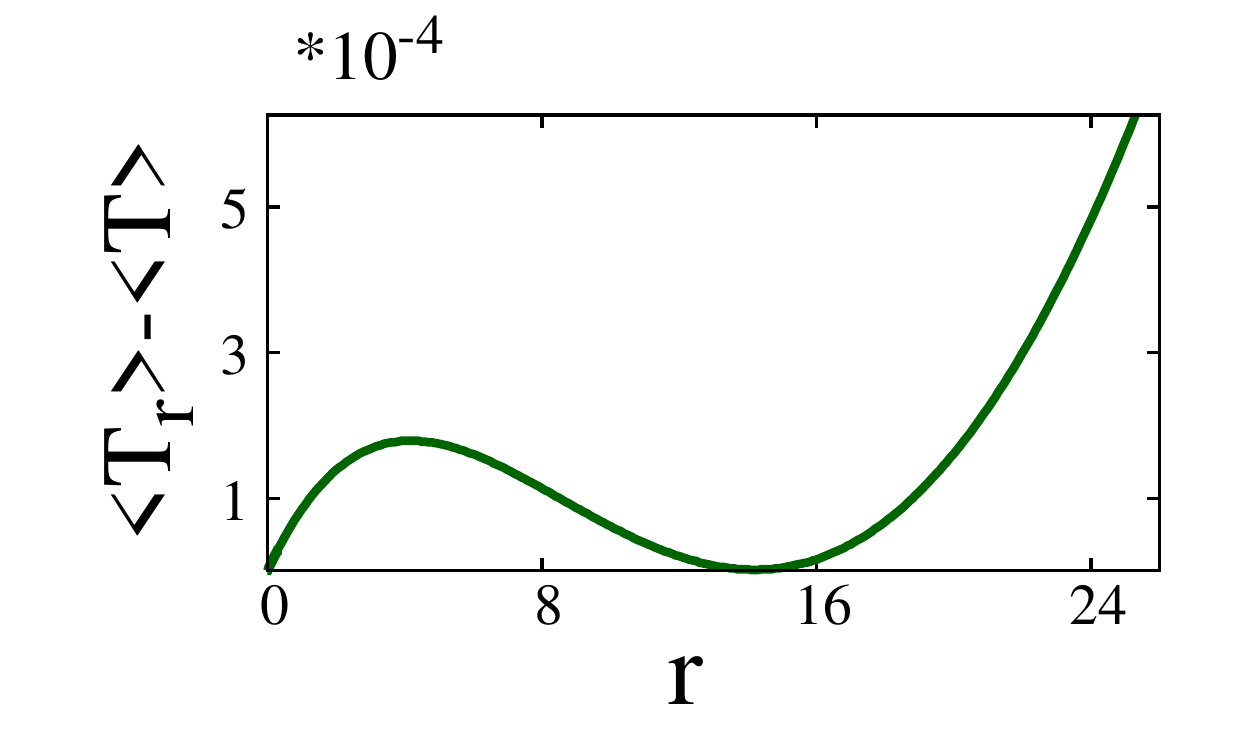}
    \caption{We plot the difference of MFPT $\langle T_r \rangle$ with resetting and MFPT $\langle T \rangle$ without resetting, as a function of $r$, in $d=2$. The double minima at $r_*=0$ to $r_*=14.2 $ show the discontinuous jump in ORR. The parameters at the transition point are given by $D=1$, $a =0.1$, $R_m=1.1$, $\langle T_{on} \rangle=0.005$ and $k=4.0$.}
    \label{fig:Model3_MFPT_vs_r}
\end{figure}

\begin{figure}[ht!]
  \begin{subfigure}{0.475\textwidth}
  \includegraphics[width = 1\textwidth,height=0.23\textheight]{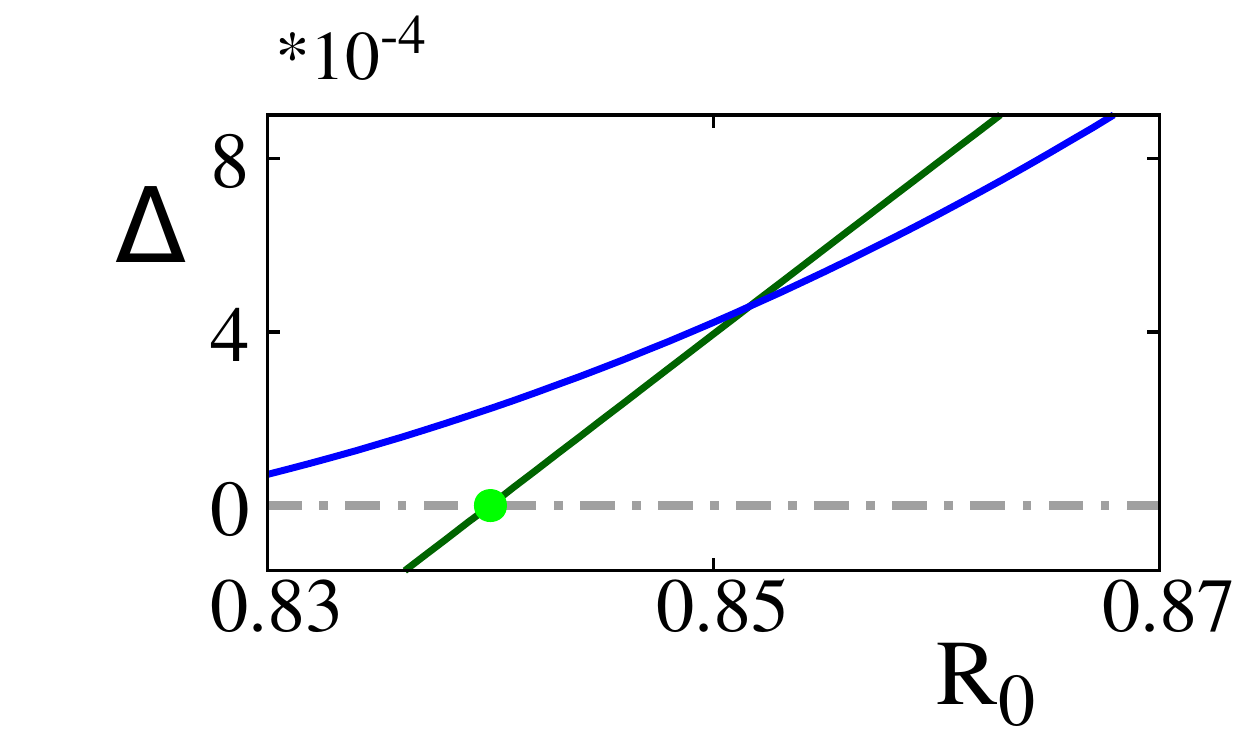}
    \caption[]{} 
    \label{fig:2nd_1d_dim_linear_pot}
   \end{subfigure}
 \begin{subfigure}{0.475\textwidth}
     \includegraphics[width = 1\textwidth,height=0.23\textheight]{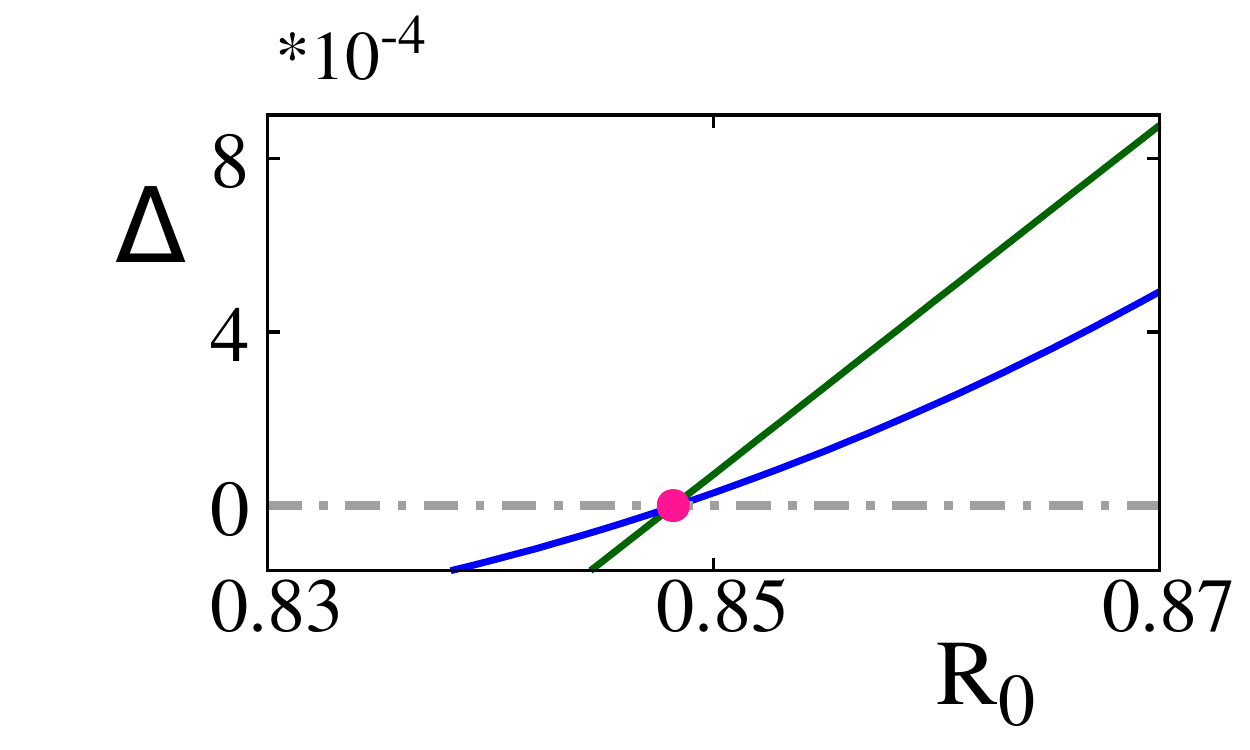}
     \caption[]{} 
    \label{fig:TCP_d_dim_pw_linear_ordr}
  \end{subfigure}
 \caption{We demonstrate the use of Eqs. (6) and (7) here to obtain the critical and tri-critical points. We plot  $\Delta$ vs. $R_0$ for a given $k$, where  $\Delta\equiv\Delta_2=\sigma^2-\langle T\rangle^2 -2\langle T\rangle \langle T_{on}\rangle$  (green lines) and $\Delta\equiv\Delta_3 = \langle T^3\rangle-6\langle T\rangle(\sigma^2+\langle T_{on}\rangle^2)$ (blue lines).  In (a) we see that at the critical point $k = k_c=0.6$ and $R_{0c}=0.84$ (green circle),  $\Delta_2 = 0$ but $\Delta_3 > 0$ (see Eq. (\ref{Conti_Trans})). In (b) we see that at the tri-critical point  $k^{2t}=0.88$ and $R_{2c}=0.848$ (pink circle), both $\Delta_2$ and $\Delta_3$ crosses zero simultaneously (see Eq. (\ref{Trans_TCP})). The parameters are $D=1$, $a=0.1$, $L=1.1$ and $d=1$.
 }
   \label{fig:1_d_cont_TCP_Ton_0}
 \end{figure}
\newpage
\section*{References}
\bibliography{Revised_J_Phys_A_V2}

\providecommand{\newblock}{}
\begin{thebibliography}{10}
\expandafter\ifx\csname url\endcsname\relax
  \def\url#1{{\tt #1}}\fi
\expandafter\ifx\csname urlprefix\endcsname\relax\def\urlprefix{URL }\fi
\providecommand{\eprint}[2][]{\url{#2}}

\bibitem{PRL_Evan_Staya_2011}
Evans M~R and Majumdar S~N 2011 {\em Phys. Rev. Lett.\/} {\bf 106}(16) 160601

\bibitem{J_Phys_A_Evan_Staya_2011Position}
Evans M~R and Majumdar S~N 2011 {\em J. Phys. A: Math. and Theor.\/} {\bf 44}
  435001

\bibitem{J_Phys_A_Evans_Non_equil_2013}
Evans M~R, Majumdar S~N and Mallick K 2013 {\em J. Phys. A: Math. and Theor.\/}
  {\bf 46} 185001

\bibitem{J_Phys_A_math_cogu_diff_w_r_Durang_2014}
Durang X, Henkel M and Park H 2014 {\em J. Phys. A: Math. and Theor.\/} {\bf
  47} 045002

\bibitem{Ref_1_PRL_Satya_Sanjib_First_2014}
Kusmierz L, Majumdar S~N, Sabhapandit S and Schehr G 2014 {\em Phys. Rev.
  Lett.\/} {\bf 113}(22) 220602

\bibitem{J_Phys_A_Evans_Staya_higher_d_2014}
Evans M~R and Majumdar S~N 2014 {\em J. Phys. A: Math. and Theor.\/} {\bf 47}
  285001

\bibitem{J_Stat_Mech_determinitic_Bhat_2016}
Bhat U, Bacco C~D and Redner S 2016 {\em J. Stat. Mech.: Theor. and Exp.\/}
  {\bf 2016} 083401

\bibitem{PRE_pwer_law_rate_Apoorva_ShamikG_2016}
Nagar A and Gupta S 2016 {\em Phys. Rev. E\/} {\bf 93}(6) 060102

\bibitem{J_Phys_A_ApalKundu_Evan2016reset_t}
Pal A, Kundu A and Evans M~R 2016 {\em J. Phys. A\/} {\bf 49} 225001

\bibitem{PRL_Staya_fluctuating_interface_2014}
Gupta S, Majumdar S~N and Schehr G 2014 {\em Phys. Rev. Lett.\/} {\bf 112}(22)
  220601

\bibitem{Arxiv_RK_singh_Probability_Relax}
Singh R~K, G\'orska K and Sandev T 2022 {\em Phys. Rev. E\/} {\bf 105}(6)
  064133

\bibitem{Review_Evans_Satya_Reset_application_2019}
Evans M~R, Majumdar S~N and Schehr G {\em J. Phys. A: Math. and Theor.\/}

\bibitem{PRL_Reuveni_2016First}
Reuveni S 2016 {\em Phys. Rev. Lett.\/} {\bf 116}(17) 170601

\bibitem{PRL_Montanari_2002}
Montanari A and Zecchina R 2002 {\em Phys. Rev. Lett.\/} {\bf 88}(17) 178701

\bibitem{PRE_Apal_descrete_time_renewal_2021}
Bonomo O~L and Pal A 2021 {\em Phys. Rev. E\/} {\bf 103}(5) 052129

\bibitem{PNAS_Reuveni_Shlomi_2014_MicMenten}
Reuveni S, Urbakh M and Klafter J 2014 {\em Proc. Natl. Acad. Sci.\/} {\bf 111}
  4391--4396

\bibitem{PRE_Sajib_2015Morkov_reset}
Meylahn J~M, Sabhapandit S and Touchette H 2015 {\em Phys. Rev. E\/} {\bf
  92}(6) 062148

\bibitem{Ref_2_PRE_Lukas_Ewa_2015}
Ku\ifmmode~\acute{s}\else \'{s}\fi{}mierz L and Gudowska-Nowak E 2015 {\em
  Phys. Rev. E\/} {\bf 92}(5) 052127

\bibitem{Ref_3_PRL_Belan_2018}
Belan S 2018 {\em Phys. Rev. Lett.\/} {\bf 120}(8) 080601

\bibitem{PRR_Ising_Model_w_reset_Majumdar_2020}
Magoni M, Majumdar S~N and Schehr G 2020 {\em Phys. Rev. Research\/} {\bf 2}(3)
  033182

\bibitem{Ref_4_PRR_Reuveni_home_return_2020}
Pal A, Ku\ifmmode~\acute{s}\else \'{s}\fi{}mierz L and Reuveni S 2020 {\em
  Phys. Rev. Research\/} {\bf 2}(4) 043174

\bibitem{PhysRevE_Bressloff_2020}
Bressloff P~C 2020 {\em Phys. Rev. E\/} {\bf 102}(2) 022134

\bibitem{PRE_Staya_Sanjib_2015_temporal_relaxation}
Majumdar S~N, Sabhapandit S and Schehr G 2015 {\em Phys. Rev. E\/} {\bf 91}(5)
  052131

\bibitem{J_Phys_A_accele_reset_Singh_2020}
Singh P 2020 {\em J. Phys. A: Math. and Theor.\/} {\bf 53} 405005

\bibitem{PRE_Bodrova_D(t)_reset_2019}
Bodrova A~S, Chechkin A~V and Sokolov I~M 2019 {\em Phys. Rev. E\/} {\bf
  100}(1) 012119

\bibitem{PRE_Telegrap_w_reset_2018}
Masoliver J 2019 {\em Phys. Rev. E\/} {\bf 99}(1) 012121

\bibitem{PRL_Optimization_w_reset_FP_2020}
De~Bruyne B, Randon-Furling J and Redner S 2020 {\em Phys. Rev. Lett.\/} {\bf
  125}(5) 050602

\bibitem{PRR_Madian_mode_FP_w_reset_2020}
Belan S 2020 {\em Phys. Rev. Research\/} {\bf 2}(1) 013243

\bibitem{PRE_Extereme_Statistics_w_reset_2021}
Singh P and Pal A 2021 {\em Phys. Rev. E\/} {\bf 103}(5) 052119

\bibitem{PRE_Non_inst_reset_Badrova_2020}
Bodrova A~S and Sokolov I~M 2020 {\em Phys. Rev. E\/} {\bf 101}(5) 052130

\bibitem{PRE_Non_inst_reset_Pal_Ku_Reuveni_2019}
Pal A, Ku\ifmmode~\acute{s}\else \'{s}\fi{}mierz L and Reuveni S 2019 {\em
  Phys. Rev. E\/} {\bf 100}(4) 040101

\bibitem{New_Jour_Phys_Non_inst_reset_Pal_2019}
Pal A, Łukasz Kuśmierz and Reuveni S 2019 {\em New Journal of Physics\/} {\bf
  21} 113024

\bibitem{PRE_Asymmetric_reset_D_Gupta_2020}
Plata C~A, Gupta D and Azaele S 2020 {\em Phys. Rev. E\/} {\bf 102}(5) 052116

\bibitem{J_Phys_Commun_Susceptivility_reset_Grange_2020}
Grange P 2020 {\em Journal of Physics Communications\/} {\bf 4} 095018

\bibitem{J_Phys_A_non_inst_reset_linear_pot_Gupta_2021}
Gupta D, Pal A and Kundu A 2021 {\em J. Stat. Mech.: Theor. and Exp.\/} {\bf
  2021} 043202

\bibitem{J_Stat_Mech_Reset_under_damped_Gupta_2019}
Gupta D 2019 {\em J. Stat. Mech.: Theor. and Exp.\/} {\bf 2019} 033212

\bibitem{Soft_matter_Reset_w_Lorentz_force_2021}
Abdoli I and Sharma A 2021 {\em Soft Matter\/} {\bf 17}(5) 1307--1316

\bibitem{PRL_Reset_w_branching_Reuveni_2020}
Pal A, Eliazar I and Reuveni S 2019 {\em Phys. Rev. Lett.\/} {\bf 122}(2)
  020602

\bibitem{J_Chem_letter_Reuveni_experimental-reset_2020}
Tal-Friedman O, Pal A, Sekhon A, Reuveni S and Roichman Y 2020 {\em J. Phys.
  Chem. Lett.\/} {\bf 11} 7350--5

\bibitem{PRR_Experiment_evidence_Majumdar_2020}
Besga B, Bovon A, Petrosyan A, Majumdar S~N and Ciliberto S 2020 {\em Phys.
  Rev. Research\/} {\bf 2}(3) 032029

\bibitem{Brief_Review_reset_Shamik_G_2022}
Gupta S and Jayannavar A~M 2022 {\em Frontiers in Physics\/} {\bf 10}

\bibitem{PRE_backtrac_RNA_polymer_Roldan_2016}
Rold\'an E, Lisica A, S\'anchez-Taltavull D and Grill S~W 2016 {\em Phys. Rev.
  E\/} {\bf 93}(6) 062411

\bibitem{J_Phys_A_Run_and_tumble_Evans_Staya_2018}
Evans M~R and Majumdar S~N 2018 {\em J. Phys. A: Math. and Theor.\/} {\bf 51}
  475003

\bibitem{J_Phys_A_Run_and_tumble_w_reset_2d_Santra_2020}
Santra I, Basu U and Sabhapandit S 2020 {\em J. Stat. Mech.: Theor. and Exp.\/}
  {\bf 2020} 113206

\bibitem{PRE_mutiple_target_Bressloff_2020}
Bressloff P~C 2020 {\em Phys. Rev. E\/} {\bf 102}(2) 022115

\bibitem{EPL_Antiviral_reset_Ramoso_2020}
Ramoso A~M, Magalang J~A, S{\'{a}}nchez-Taltavull D, Esguerra J~P and
  Rold{\'{a}}n {\'{E}} 2020 {\em Europhysics Letters\/} {\bf 132} 50003

\bibitem{J_Phys_A_partial_ab_Bresloff_Schumm_2021}
Schumm R~D and Bressloff P~C 2021 {\em J. Phys. A: Math. and Theor.\/} {\bf 54}
  404004

\bibitem{PRE_Reuveni_Shlomi2015_MicMenten}
Rotbart T, Reuveni S and Urbakh M 2015 {\em Phys. Rev. E\/} {\bf 92}(6) 060101

\bibitem{Nature_Single_Enzyme_Robin_S_Reuveni_2018}
Robin T, Reuveni S and Urbakh M 2018 {\em Nature Communications\/} {\bf 9}
  2041--1723

\bibitem{PRE_Optimal_reset_polination_2019}
Robin T, Hadany L and Urbakh M 2019 {\em Phys. Rev. E\/} {\bf 99}(5) 052119

\bibitem{Compt_sys_Optimal_reset_Lorenz_2021}
Lorenz J 2021 {\em Comput. Syst.\/} {\bf 65} 1143--64

\bibitem{PRL_APAl_Reuveni_2017FirstR}
Pal A and Reuveni S 2017 {\em Phys. Rev. Lett.\/} {\bf 118}(3) 030603

\bibitem{PRE_APal_2015Potential}
Pal A 2015 {\em Phys. Rev. E\/} {\bf 91}(1) 012113

\bibitem{PRE_Saeed_2019}
Ahmad S, Nayak I, Bansal A, Nandi A and Das D 2019 {\em Phys. Rev. E\/} {\bf
  99}(2) 022130

\bibitem{J_Phys_A_Ray_Debasish_Reuveni_2019}
Ray S, Mondal D and Reuveni S 2019 {\em Journal of Physics A: Mathematical and
  Theoretical\/} {\bf 52} 255002

\bibitem{PRR_Pal_Parsad_Landau_2019}
Pal A and Prasad V~V 2019 {\em Phys. Rev. Research\/} {\bf 1}(3) 032001

\bibitem{J_Chem_Phys_Reuveni_log_Pot_2020}
Ray S and Reuveni S 2020 {\em The Journal of Chemical Physics\/} {\bf 152}
  234110

\bibitem{PRE_Saeed_Das_2020}
Ahmad S and Das D 2020 {\em Phys. Rev. E\/} {\bf 102}(3) 032145

\bibitem{J_Chem_Phys_Reuveni_Ray_dble_w_2021}
Ray S and Reuveni S 2021 {\em The Journal of Chemical Physics\/} {\bf 154}
  171103

\bibitem{PRE_Huang_higher_d_reset_2_ab_2022}
Chen H and Huang F 2022 {\em Phys. Rev. E\/} {\bf 105}(3) 034109

\bibitem{PRE_Saeed_2022}
Ahmad S, Rijal K and Das D 2022 {\em Phys. Rev. E\/} {\bf 105}(4) 044134

\bibitem{J_Phys_A_Christo_Reset_Boun_2015}
Christou C and Schadschneider A 2015 {\em J. Phys. A\/} {\bf 48} 285003

\bibitem{PRE_Cristou_reset_circle_2018}
Chatterjee A, Christou C and Schadschneider A 2018 {\em Phys. Rev. E\/} {\bf
  97}(6) 062106

\bibitem{Ref_5_PRE_2015}
Campos D and M\'endez V 2015 {\em Phys. Rev. E\/} {\bf 92}(6) 062115

\bibitem{J_Phys_A_Refractory_Period_w_reset_Evans_2019}
Evans M~R and Majumdar S~N 2018 {\em J. Phys. A.: Math. and Theor.\/} {\bf 52}
  01LT01

\bibitem{J_Phys_A_reset_confining_pot_Metzler_Singh_2020}
Singh R~K, Metzler R and Sandev T 2020 {\em Journal of Physics A: Mathematical
  and Theoretical\/} {\bf 53} 505003

\bibitem{Mthemetica_HypergeometricU_Series}
\url{https://functions.wolfram.com/HypergeometricFunctions/HypergeometricU/06/01/03/02}

\bibitem{simmons2016differential}
Simmons G~F 2016 {\em Differential equations with applications and historical
  notes\/} (CRC Press)

\bibitem{Arfken_7ed_book}
Arfken G~B 2013 {\em Mathematical Method for Physicists\/} 7th ed (Academic
  Press) ISBN 9780123846549

\bibitem{gard_book_2004}
Gardiner C 2004 {\em Handbook of Stochastic Methods for Physics, Chemistry, and
  the Natural Sciences\/} Springer complexity (Springer) ISBN 9783540208822

\bibitem{redner2001guide}
Redner S 2001 {\em A Guide to First-Passage Processes\/} (Cambridge University
  Press) ISBN 9780521652483

\end{thebibliography}

\end{document}